\documentclass[aps,prd,reprint,amsmath,amssymb,groupedaddress,nofootinbib,floatfix]{revtex4-1} 
\usepackage{subfig}
\usepackage{graphicx}

\begin{document}
\title{Improved Semileptonic Form Factor Calculations in Lattice QCD}
\author{Richard Evans}\email{richard.evans@physik.uni-regensburg.de}\affiliation{Institut f\"{u}r Theoretische Physik, Universit\"{a}t Regensburg, 93040 Regensburg, Germany}\author{Gunnar Bali}\email{gunnar.bali@physik.uni-regensburg.de}\author{Sara Collins}\email{sara.collins@physik.uni-regensburg.de}
 \affiliation{Institut f\"{u}r Theoretische Physik, Universit\"{a}t Regensburg, 93040 Regensburg, Germany}
\collaboration{QCDSF Collaboration}
\noaffiliation
\date{\today}
\begin{abstract}
We investigate the computational efficiency of two stochastic based alternatives to the Sequential Propagator Method used in Lattice QCD calculations of
heavy-light semileptonic form factors.  In the first method, we replace the sequential propagator, which couples the calculation of two of the three propagators required for the calculation, with a stochastic propagator so that the calculations of all three propagators are independent.  This method is more flexible than the Sequential Propagator Method but introduces stochastic noise.  We study the noise to determine when this method becomes competitive with the Sequential Propagator Method, and find that for any practical calculation it is competitive with or superior to the Sequential Propagator Method.  We also examine a second stochastic method, the so-called ``one-end trick", concluding it is relatively inefficient in this context.

The investigation is carried out on two gauge field ensembles, using the non-perturbatively
improved Wilson-Sheikholeslami-Wohlert action with $N_f=2$ mass-degenerate sea
quarks.  The two ensembles have similar lattice spacings but different sea quark masses.  We use the first stochastic method to extract ${\mathcal O}(a)$-improved, matched lattice results for the semileptonic form factors on the ensemble with lighter sea quarks, extracting $f_+(0)$.  
\end{abstract}

\maketitle

\section{Introduction}
Experimental measurements of heavy-light semileptonic decays, combined with theoretical input,  can be used to extract the Cabibbo-Kobayashi-Maskawa (CKM) matrix elements $|V_{ub}|$, $|V_{cb}|$, $|V_{cd}|$, and $|V_{cs}|$.  The determination of these matrix elements provides constraints on the CKM Unitarity Triangle and thus tests the Standard Model.  Conversely, $|V_{cd}|$ and $|V_{cs}|$ are known with high precision, and can be used to test the lattice techniques involved in calculating the decay rates of $D$ mesons.  Recent progress reports of current $D$ meson semileptonic decay calculations are presented in Refs.~\cite{Bailey:2009pz,Na:2009au,DiVita:2009by}.  The most recent study where systematic errors are taken into account was presented in Ref.~\cite{Aubin:2004ej}.  In this paper we study the efficiency of two stochastic propagator based alternatives to the Sequential Propagator Method used in all of the above calculations.

 The calculation we focus on is the semileptonic decay of a heavy-light pseudoscalar meson ($H=D,D_s$) to a light-light pseudoscalar meson ($P=\pi,K$).  For processes $H\to P$ the differential decay rate can be parameterized as a product of CKM matrix elements, perturbatively known quantities, and poorly known non-perturbative quantities, 
\begin{align}
\frac {d \Gamma}{dq^2}=\underbrace{|V_{cl}|^2}_{\textrm{CKM}}\underbrace{\frac{G_F^2}{192\pi^2m^3_{H}}\lambda^{3/2}(q^2)}_{\textrm{perturbatively known}}\underbrace{|f_{+}(q^2)|^2}_{\textrm{form factor}}\,,
\end{align}
where $q^2=(p_H-p_P)^2$ is the squared difference between the initial and final state four-momenta and  $l\in \left\{d,s\right\}$.  The greatest source of uncertainty in the theoretical calculation of the decay rate is due to the non-perturbative interactions parameterized by the form factor $f_+(q^2)$.

\begin{figure}
\includegraphics[width=0.95\linewidth,clip]{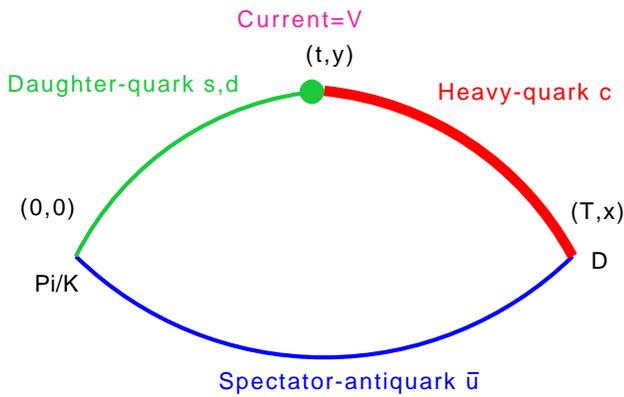}
\caption{The three-point function, Eq.~(\protect\ref{eq:c3}).  Point-to-all propagators are used for the daughter-quark and spectator-antiquark.  The heavy-quark propagator is constructed either by sequential
or by stochastic all-to-all
propagator techniques.
 \label{fig:c3}}
\end{figure}

  These interactions appear in the hadronic matrix element $\langle H(p_H)|V_{\mu}(q^2)| P(p_P)\rangle$, where $V_{\mu}=\bar{\psi_c}\gamma_{\mu}\psi_l$ is a weak flavor-changing vector current and $\psi_c$ is the charm heavy-quark and $\psi_l$ is the $d$ or $s$ daughter-quark spinor.
 The matrix element can be parameterized as a linear combination of the vector, $f_+$, and scalar, $f_0$, form factors,
\begin{align}\begin{split}
\langle H(p_H)&|V_{\mu}(q^2)| P(p_P)\rangle=\\&\left\{ p_H+p_P-q(m_H^2-m_P^2)/q^2\right\}_{\mu}f_+(q^2)\\&+\left\{ q(m_H^2-m_P^2)/q^2\right\}_{\mu} f_0(q^2)\,.\end{split}
\end{align}  
On the lattice the matrix element is extracted from three-point functions of the following form,
\begin{align}\label{eq:c3}\begin{split}
C_3(T,t&;\mathbf{p}_H,\mathbf{q})=\sum_{\mathbf{x},\mathbf{y}}e^{-i\mathbf{p}_H\cdot \mathbf{x}}e^{\mathbf{q}\cdot\mathbf{y}}\\&\times
\langle 0|\bar{\psi}_u\gamma_5\psi_c(\mathbf{x},T)\bar{\psi}_c\gamma_{\mu} \psi_l(\mathbf{y},t) \bar{\psi}_l \gamma_5 \psi_u(\mathbf{0},0)|0\rangle\\
=&-\sum_{\mathbf{x},\mathbf{y}}e^{-i\mathbf{p}_H\cdot \mathbf{x}}e^{\mathbf{q}\cdot\mathbf{y}}
\,\mathrm{Tr}\left\langle M^{-1}_u(\mathbf{0},0;\mathbf{x},T))\right.\\&\times\left.\gamma_5 M^{-1}_c(\mathbf{x},T;\mathbf{y},t)\gamma_{\mu}M_l^{-1}(\mathbf{y},t;\mathbf{0},0)\gamma_5\right\rangle\,,
\end{split}\end{align}
where $\bar{\psi}_u$ is the spectator-antiquark spinor,  $M^{-1}_q$ is the propagator for quark flavor $q$, and $T$ is the time-slice at which the sink is fixed. The three-point function is shown schematically in Fig.~\ref{fig:c3}. In the limit of large time separations Eq.~(\ref{eq:c3}) has the form,
\begin{align}\begin{split}
\mathop {\lim} \limits_{T\gg t \gg0}&C_3(T,t;\mathbf{p}_H,q)=\\ &\frac {Z_P}{2E_P} \frac{ Z_H}{2E_H} \langle H(p_H)|V_{\mu}|P(p_P)\rangle  e^{-E_P t} e^{-E_H(T-t)}.
\label{eq:c3tlim}\end{split}
\end{align}
The wave-function amplitudes, $Z_P$ and $Z_H$, and energies can be removed by forming ratios with or performing simultaneous fits to the $P$ and $H$ meson two-point functions,
\begin{equation}
C_2^{P}(t)\stackrel{t\gg 0}{\longrightarrow} \frac {|Z_{P}|^2} {2E_{P}}e^{-E_{P}t}\,,  \quad C_2^{H}(t)\stackrel{t\gg 0}{\longrightarrow} \frac {|Z_{H}|^2} {2E_{H}}e^{-E_{H}t}\,.
\end{equation}

Currently, the approach of most lattice calculations of  semileptonic decay rates of $D$ mesons involves constructing the three-point functions using a so-called sequential (or extended) propagator, the Sequential Propagator Method \cite{Martinelli:1988rr}.  As is shown in Eq.~(\ref{eq:c3}), this method uses two light-quark, point-to-all propagators $M_u^{-1}$ and $M_l^{-1}$, which may or may not be mass degenerate.  It is routinely the case that these computationally expensive (expensive relative to the heavy-quark propagator) light-quark propagators have been previously generated to study other lattice quantities, and then stored.  For example, the pseudoscalar pion mass is required to guide chiral extrapolations of many lattice quantities.

  To construct the desired three-point functions, the light-quark propagators are combined via the Sequential Propagator Method with the heavy-quark propagator generated ``on-the-fly".  The point-to-all light-quark propagators connect the source, at $(0,\mathbf{0})$, to the current at $(t,\mathbf{y})$ and sink at $(T,\mathbf{x})$.  A point source is sufficient because of translational invariance.  The heavy-quark propagator, $M_c^{-1}(\mathbf{x},T;\mathbf{y},t)=\gamma_5 [M_c^{-1}(\mathbf{y},t;\mathbf{x},T)]^{\dagger} \gamma_5 $ in Eq.~(\ref{eq:c3}), must then connect all lattice sites at the vector current time-slice to all lattice sites at the sink time-slice. 
  This is achieved by solving the heavy-quark action on the spectator-antiquark propagator sink at a given time-slice $T$, with a momentum and $\Gamma$ matrix insertion. Although inverting the heavy-quark action with a $\kappa$ corresponding to the charm quark mass is inexpensive, relative to the calculation of the light-quark propagators, this approach requires an additional 12 heavy-quark solves for every spectator-quark, $D$ meson momentum, and/or $D$ meson smearing.    If several combinations of these parameters are wanted, the cost of calculating the heavy-quark propagators becomes significant.    In particular, at small lattice spacings the momenta are more finely grained, and therefore more three-point functions will have an appreciable signal, requiring many more heavy-quark solves if all data are to be utilized. 
 
Another approach is the One-end Method.  This method combines sequential propagator techniques with stochastic propagators,   through the so-called ``one-end trick" as described for two-point functions in Refs.~\cite{Foster:1998vw,McNeile:2002fh}.  The method has been used successfully in form factor calculations in light-light systems \cite{Simula:2007fa,Boyle:2008yd}.  This method provides an additional volume average and momentum projection which may greatly reduce the gauge noise in the correlators. However, additional propagators are required for every source and sink momentum combination, suggesting it is computationally expensive relative to the Sequential Propagator Method.  Nevertheless, there is the possibility that the three-point function noise is so greatly reduced that very few momentum points would be required in the $q^2\to 0$ extrapolation and a gain in the overall efficiency is realized.  

A more flexible method than the previous two that we also investigate in this paper uses an all-to-all \cite{Bitar:1988bb}, stochastically generated, propagator to connect the sink to the current.    We will refer to this approach as the Stochastic Propagator Method. With this method all momenta and sink smearings are in principle available using only a single spin-color inversion.  In addition, no new heavy-quark inversions would be required to calculate the three-point function using multiple (previously generated) spectator-quark propagators on the same configuration, as might be useful for chiral extrapolations in the valence light-quark mass.  There is also the interesting possibility that correlators could be generated with multiple current and sink time slices with no additional inversions. 
 As we will show, the stochastic noise is significant and must be reduced by averaging over multiple noise vectors and partitioning \cite{Bernardson:1993yg,Viehoff:1997wi,Wilcox:1999ab,Foley:2005ac,Ehmann:2009ki}.  In particular, time partitioning is required for efficient reduction of noise, even when using other variance reduction techniques such as the hopping parameter acceleration (HPA) \cite{Thron:1997iy,Michael:1999rs,Bali:2005pb,Bali:2005fu}.  
    
 In the following, we study and minimize the stochastic noise appearing in the methods using variance reduction techniques. Comparing the errors and costs of the three methods in Sec.~\ref{comps}, we determine the One-end Method to be inefficient but the Stochastic Propagator Method to be promising.  We discuss when the Stochastic Propagator Method is more efficient than the Sequential Propagator Method, and finally apply the method to determine $f_+(0)$ on a single ensemble.  This study is an extension of our preliminary results presented in Ref.~\cite{Evans:2009xp}.

   \section{\label{sec:gsu}Calculational details}
\begin{table}
\caption{Simulation Parameters for the 2 QCDSF ensembles $\mathcal H$ and $\mathcal L$ used.   The scale was set from $r_0=0.467$ fm, which was obtained by chirally extrapolating the nucleon mass times $r_0$, $m_Nr_0$,  to the physical point at $\beta=5.29$ \cite{Gockeler:2008we}.
\label{tab:params}}
\begin{ruledtabular}
\begin{tabular}{ccc}
parameter& ensemble heavy ($\mathcal H$)  & ensemble light ($\mathcal L$) \\
\hline
$\beta$ & 5.29 & 5.29 \\
Volume & $16^3\times 32$ & $24^3\times 48$ \\
$\kappa_{\mathrm{sea}}$ &   0.13500    &  0.13632   \\
$\kappa_{\mathrm{light, \,valence}}$ & 0.13500 & 0.13609 \\
$\kappa_{\mathrm{heavy,\, valence}}$ & 0.11628 & 0.12444 \\
\hline
$a$ & 0.089 fm & 0.076 fm\\
$m_{\pi,\mathrm{sea}}$ & $\sim 930$ MeV & $\sim 290$ MeV\\
$m_{\pi,\mathrm{valence}}$  &$\sim 930$ MeV & $\sim 430$ MeV\\
$m_{D}$ & $\sim 2.43$ GeV& $\sim 1.93$ GeV\\ 
\end{tabular}
\end{ruledtabular}
\end{table}

Two QCDSF ensembles with two degenerate sea quark flavors  \cite{Ali Khan:2003cu} were used in this study, with the parameters shown in Tab.~\ref{tab:params}. The ensembles were generated using the Wilson plaquette action for the gluons and the non-perturbatively improved  Sheikholeslami-Wohlert action for the fermions.  Both actions have errors starting at ${\mathcal O}(a^2)$.  We refer to the ensemble with relatively heavy sea-quarks as $\mathcal H$, and to the one with relatively light sea-quarks as $\mathcal L$. Wuppertal smearing  \cite{Gusken:1989ad} on top of APE smeared gauge links is performed on the interpolating fields in the two- and three-point functions, using smearing parameters that optimize the overlap with the light-light meson ground state. 

 The ensemble with the heavier sea-quark mass, $\mathcal H$, is significantly cheaper to work with, and so we performed some tests on it exclusively.
  The main results in this paper will, however, focus on the ensemble with the lighter sea-quark mass, $\mathcal L$, since these parameters are much closer to the physical limit and the behavior of the noise should be influenced by the particle masses, lattice volume etc.  As it turned out, the three-point functions generated on ensemble $\mathcal H$ had a noise behavior very similar to that on ensemble $\mathcal L$, suggesting the influence of these simulation parameters on the noise is not large.  
  
\section{\label{sec:mvar} Sequential and Stochastic Methods}
\subsection{Sequential Propagator Method}
As stated in the introduction, the standard method for calculating $D$ meson semileptonic three-point functions is the Sequential Propagator Method.  The sequential propagator provides a way to calculate a heavy-quark propagator that connects all spatial sites $\mathbf{x}$ at the sink time-slice $T$, to all sites $\mathbf{y}$ and $t$ at the vector current insertion: the propagator $M_c^{-1}$ that appears in Eq.~(\ref{eq:c3}).  Algorithmically this amounts to taking a single time-slice $T$ of the spectator-quark propagator, $M^{-1}_u(\mathbf{x},T;\mathbf{0},0)$, and projecting it onto the sink momentum $\mathbf{p}_H$, the sink smearing $W_{\mathrm{snk}}$, and the sink $\Gamma$ matrix. Finally, the heavy-quark Dirac operator is inverted on this ``Sequential source" to get the sequential propagator,
 \begin{align}\begin{split}   
G&(\mathbf{y},t;\mathbf{p}_H,T;\mathbf{0},0)=\\&\sum_{\mathbf{x}}M^{-1}_c(\mathbf{y},t;\mathbf{x},T) \underbrace{ \Gamma W e^{i\mathbf{p}_H\cdot \mathbf{x}}M^{-1}_u(\mathbf{x},T;\mathbf{0},0)}_{\textrm{``sequential source"}}\,. \end{split}
\end{align}
The sequential propagator, $G(\mathbf{0},0;\mathbf{p},T;\mathbf{y},t)=\gamma_5 [G(\mathbf{y},t;\mathbf{p},T;\mathbf{0},0)]^{\dagger}\gamma_5$, can then be combined with the daughter light-quark propagator, $M^{-1}_l(\mathbf{y},t;\mathbf{0},0)$, and appropriate gamma matrices to arrive at Eq.~(\ref{eq:c3}).
This method requires 12 heavy-quark solves (one for each spin and color index) for each distinct sink momentum $\mathbf{p}_H$, sink smearing $W_{\mathrm{snk}}$, sink time $T$, spectator-quark mass, and sink $\Gamma$ matrix.  The computational effort required for this procedure can become significant if many combinations of these parameters are needed, as would be required in partially-quenched chiral extrapolations or when using the Variational Method \cite{Michael:1985ne,Luscher:1990ck} for studying excited states.

\subsection{Stochastic Propagator Method}
With this method, the sequential propagator is replaced by an all-to-all propagator, or more precisely with a time-slice-to-all ($T$-to-all) propagator\footnote{A true all-to-all propagator, where all time-slices have support, could be used.  As we discuss in Sec.~\ref{PART}, this turns out to be  inefficient.}.   We construct $T$-to-all propagators by generating noise vectors $\eta^{[\ell]}_j(\mathbf{x},T)$, $\ell=1,\ldots,N$, using complex $\mathbb{Z}_2$ noise, 
 with the property, 
\begin{equation}
 \frac 1 N \sum_{\ell} \eta^{[\ell]}_i(x) \eta^{\dagger [\ell]}_j(z)=\delta_{xz}\delta_{ij}+{\mathcal O}\left(1/\sqrt{N}\right)\,, 
 \label{eq:z2}
\end{equation}
where $x=(\mathbf{x},T)$, $z=(\mathbf{z},T)$, and $i,j$ are combined spin and color indices.  We then solve the linear system with the charm quark Dirac operator $M_{c,kj}(x,y)$ and source vector $\eta_j^{[\ell]}(\mathbf{x},T)$, for the solution vector $\psi_j^{[\ell]}(\mathbf{y},t)$:
\begin{align}\begin{split}
&\sum_{j,\mathbf{y},t} M_{c,kj}(\mathbf{z},T;\mathbf{y},t)\psi^{[\ell]}_j(\mathbf{y},t)=\eta^{[\ell]}_k(\mathbf{z},T)\\&\Rightarrow \psi^{[\ell]}_j(\mathbf{y},t)=\sum_{k,\mathbf{z}} M^{-1}_{c,jk}(\mathbf{y},t;\mathbf{z},T)\eta^{[\ell]}_k(\mathbf{z},T)\,.\end{split}
\end{align}
The average of the product of the source and solution vectors provides an unbiased estimate of the $T$-to-all heavy-quark propagator,
\begin{align}\begin{split}
\frac 1 N \sum_{\ell}& \psi^{[\ell]}_j(\mathbf{y},t) \eta^{\dagger [\ell]}_i(\mathbf{x},T)=\underbrace{ M^{-1}_{c,ji}(\mathbf{y},t;\mathbf{x},T) }_{ T\textrm{-to-all} }\\+ & \sum_{k,\mathbf{z}}M^{-1}_{c,jk}(\mathbf{y},t;\mathbf{z},T)\\
&\times\underbrace{ \left(\frac 1 N \sum_{\ell} \eta^{[\ell]}_k(\mathbf{z},T)\eta^{\dagger [\ell]}_i(\mathbf{x},T)-\delta_{\mathbf{z}\mathbf{x}}\delta_{ki}  \right)}_{ \mathbb{Z}_2\, \textrm{noise}\, \propto\,{\mathcal O}(1/\sqrt{N})}\,,\end{split}
\end{align}
where the stochastic error decreases with $N$, the number of noise vectors used.  In the limit $N\to \infty$, the exact $T$-to-all propagator is obtained. 

A stochastic estimate of Eq.~(\ref{eq:c3}) can be constructed by combining the spectator-quark and daughter-quark propagators and the $T$-to-all heavy-quark sources and solutions in the following manner,
\begin{align}\begin{split}
C_3&(T,t;\mathbf{p}_H,\mathbf{q})
=\\&-\frac 1 N \sum_{\ell,\mathbf{x},\mathbf{y}}
\mathrm{Tr}\left\langle e^{-i\mathbf{p}_H\cdot \mathbf{x}}  \Gamma_i M_u^{-1}(\mathbf{0},0;\mathbf{x},T)\Gamma_f \eta^{[\ell]}(\mathbf{x},T)\right.\\ &\qquad\qquad\times
e^{\mathbf{q}\cdot\mathbf{y}} \psi^{\dagger}(\mathbf{y},t)\Gamma M^{-1}_l(\mathbf{y},t;\mathbf{0},0) \Big\rangle\\
&  + {\mathcal O}(1/\sqrt{N})\,\end{split}
\end{align}
where the appropriate propagator smearings must be applied and $\Gamma_f=\Gamma_i=\gamma_5$ and $\Gamma=\gamma_{\mu}$.  Although the error in the correlators generated using the Stochastic Propagator Method will be greater than that of the exact result from the Sequential Propagator Method,  this difference is insignificant if the gauge noise dominates the stochastic noise.

\subsection{\label{oe} One-end Method}
An implementation of the One-end method for light-light three-point functions is introduced in Ref.~\cite{McNeile:2002fh}, and can be directly adapted to heavy-light three-point functions.  The approach begins with two $T$-to-all propagators generated from the same noise vectors, and then uses the sequential propagator technique to complete the three-point function.  We describe the case where the noise vectors, $\eta_k^{[\ell]}$, are used as the source at time-slice $T$ for the heavy-quark and spectator-quark solves.  

 The $D$ meson momentum,  $\mathbf{p}_H$, is first applied to the source vector and then a heavy-quark solve performed to obtain the solution vector,
\begin{equation}
\psi^{[\ell]}_{j,c}(\mathbf{y},t;\mathbf{p}_H,T)=\sum_{k,\mathbf{x}}M^{-1}_{c,jk}(\mathbf{y},t;\mathbf{x},T) e^{i\mathbf{p}_H\cdot\mathbf{x}}\eta^{[\ell]}_k(\mathbf{x},T)\,.
\end{equation}
A light-quark solve is then performed on the same source vector to get,
 \begin{equation}
\psi^{[\ell]}_{j',u}(\mathbf{y}',t')=\sum_{k',\mathbf{z}}M^{-1}_{u,j'k'}(\mathbf{y}',t';\mathbf{z},T)\eta^{[\ell]}_{k'}(\mathbf{z},T)\,.
\end{equation}
The pion momentum, $\mathbf{p}_P$, and $\gamma_5$ are then applied to the sink time-slice $t'=0$ of $\psi^{[\ell]}_{j',u}$, which is then used as the sequential source for the daughter-quark solves, resulting in the sequential propagator,
\begin{align}\begin{split}
G^{[\ell]}_j(\mathbf{y},t;\mathbf{p}_P,0;\mathbf{z},T)=&\sum_{j',k', \mathbf{y}',\mathbf{z}}M^{-1}_{l,jj'}(\mathbf{y},t;\mathbf{y}',0) \gamma_5 e^{i\mathbf{p}_P\cdot \mathbf{y}'}\\&\times M^{-1}_{u,j'k'}(\mathbf{y}',0;\mathbf{z},T)\eta^{[\ell]}_{k'}(\mathbf{z},T)\,.\end{split}
\end{align}
Forming the average of the product of $G_j^{[\ell]}$ with $\psi_{j,c}^{[\ell]\dagger}$ (only the spin and color indices are transposed) and inserting the appropriate current momentum $\mathbf{q}$ and current gamma structure $\gamma_5 \Gamma$ results in an estimate for Eq.~(\ref{eq:c3}),
\begin{align}\begin{split}
& C_3(T,t;\mathbf{p}_H,\mathbf{q})\\
&=  
\sum_{\mathbf{x},\mathbf{y},\mathbf{y'}} e^{i \mathbf{q}\cdot {y}}e^{i \mathbf{p}_P\cdot \mathbf{y}'}e^{-i\mathbf{p}_H \cdot \mathbf{x}} \mathrm{Tr} \left\langle \gamma_5 M_c^{-1}(\mathbf{x},T;\mathbf{y},t)\Gamma \right. \\& \qquad\times\left. M_u^{-1}(\mathbf{y},t;\mathbf{y}',0) \gamma_5 M_l^{-1}(\mathbf{y}',0;\mathbf{x},T)\right\rangle\\
&\approx\frac 1 N \sum_{\ell,\mathbf{y}} e^{i\mathbf{q}\cdot \mathbf{y}}  \psi_{H}^{[\ell]\dagger}(\mathbf{y},t;\mathbf{p}_H,T)\gamma_5 \Gamma G^{[\ell]}_j(\mathbf{y},t;\mathbf{p}_P,0;\mathbf{z},T)\,.\end{split}
\end{align}
The potential advantage of this method is that the additional volume average at the source (over $\mathbf{y}'$) and explicit projection onto $\mathbf{p}_H$ could reduce the stochastic and gauge noise. 

There is another possible procedure for constructing the three-point function using the One-end Method.  The $T$-to-all propagators could instead be used for both light-quarks and the sequential propagator technique for the heavy-quark.  The Sequential and Stochastic Propagator Methods also have similar alternatives, where the sequential  or stochastic propagator can be used for the daughter light-quark and point-to-all propagators for the heavy- and spectator-quarks.  However, with these methods it is obvious that the additional light-quark solves the alternative procedure requires would make for an inefficient calculation.  This is especially true in practice, where the point-to-all light-quark propagators commonly have been generated to calculate other quantities and stored.  It is, however, unclear which procedure is most efficient for the One-end Method, and the required smearings, momenta, and quark masses in the final calculation influence this decision.  It is also possible that one procedure results in significantly smaller errors than the other.  When a distinction is necessary we will refer to the procedure outlined in this section as the One-end Sink Method, and the alternative procedure as the One-end Source Method.

\section{Noise reduction techniques}

\subsection{\label{PART}Partitioning and Hopping Parameter Acceleration}
The noise reduction methods we studied are partitioning \cite{Bernardson:1993yg,Viehoff:1997wi,Wilcox:1999ab,Foley:2005ac} and Hopping Parameter Acceleration (HPA) \cite{Thron:1997iy,Bali:2005pb}.  With partitioning we decompose the noise vector sources into a sum of $M$ subspaces in space-time, spin, and/or color,
\begin{equation}
\eta^{[\ell]}=\sum_{m=1}^M \eta^{[\ell]}_m\,,
\end{equation} 
so that the sources have support only on their respective subspaces.  We then solve for the solution vectors $\psi^{[\ell]}_m$ on each subspace. Our $T$-to-all propagator is formed by first summing over the $M$ solution subspaces of a particular noise vector, then averaging over the product of the $N$ source and solution vectors,
\begin{equation}
\frac 1 N \sum_{\ell,m} \psi^{[\ell]}_{jm}(\mathbf{y},t) \eta^{\dagger [\ell]}_{im}(\mathbf{x},T) \approx M^{-1}_{ji}(\mathbf{y},t;\mathbf{x},T)\,.
\label{part}
\end{equation}
Partitioning has the effect of explicitly setting many of the terms contributing to the stochastic error to zero.  It has a computational overhead of $M$ additional inversions per noise vector, but has been seen in some previous calculations to reduce the noise faster than $\frac 1 {\sqrt{M}}$ \cite{Bernardson:1993yg,Viehoff:1997wi}.  Note that in the methods described thus far, $\eta^{[\ell]}$ had support on only a single time-slice, corresponding to full time partitioning.  This is not a strict requirement of the calculation: $\eta^{[\ell]}$ could have support on multiple or even all time-slices, allowing us to calculate the three-point function with multiple sinks.  We examine time partitioning in addition to other partitionings in the next section.

HPA also seeks to set certain stochastic error contributions to zero, in particular the stochastic error terms close to the diagonal, which, in general, have larger amplitudes.  Writing the Wilson action $M$ as,
\begin{equation}
2\kappa M=1-\kappa D\,,
\end{equation}      
and expanding the inverse in powers of $\kappa$,
\begin{equation}
M^{-1}=2\kappa \sum_{i=0}^{\infty}(kD)^i=2\kappa \sum_{i=0}^{k-1}(\kappa D)^i+(\kappa D)^k M^{-1}\,,
\end{equation}
we see that the first term on the right hand side should not contribute to the propagator for distances greater than $k$ lattice sites, because $D$ only connects nearest neighbor sites.  These terms do however contribute to the noise when $M^{-1}$ is stochastically estimated, due to the off-diagonal noise terms in Eq.~(\ref{part}).  We can explicitly remove these contributions by using $M^{-1}_{xy}=\left[ (\kappa D)^k M^{-1} \right]_{xy}$, where $| x-y |/a>k$ and $a$ is our lattice spacing.  

For three-point function calculations HPA  is potentially useful when multiple time-slices of the noise source vector have support. The stochastic error, in these cases, can be reduced by employing the HPA when the distance in Euclidean time between the sink and current are greater than $k$, $|T-t|/a>k$.  This has the effect of reducing the noise from the time-slices between $T$ and $t$.  Unfortunately, as we demonstrate in the next section, HPA does not reduce the error enough for multiple time-slice partitioning to be efficient in calculations of heavy-light semileptonic form factors .  In the case of full time partitioning the HPA has no effect, because there are no time-slices between $T$ and $t$ that have support.

\subsection{Comparison of noise reduction techniques}

Due to the many additional light-quark solves required in the One-end Method, it is far more expensive to test noise reduction techniques using it rather than using the Stochastic Propagator Method.  Therefore, we focus our efforts on minimizing the noise in the Stochastic Propagator Method, and make the assumption that the same techniques are effective in the One-end Method.  In the end we find the most efficient noise reduction technique, simple time and spin partitioning, provides a similar noise reduction in both the Stochastic Propagator and One-end Methods.   

We introduce the ratio of correlators we refer to as $V_0(q^2_{\max})$, which will be used as the starting point for comparing noise reduction techniques and the three different three-point methods, 
\begin{equation}
V_0(q^2_{\max})= \frac{C_3(T,t;\mathbf{p}=0,\mathbf{q}=0)} {C_2^{\pi}(t)C_2^{D}(T-t)}\,, 
\label{v0}
\end{equation}
where $\mathop {\lim} \limits_{T\gg t \gg0}V_0(q^2_{\max}) \to  \frac{1}{Z_{\pi}Z_D} \langle \pi(\mathbf{0})|V_{0}|D(\mathbf{0})\rangle$.
 We later form and study ratios of other correlators with different momenta and $\Gamma$ insertions, but find $V_0(q^2_{\max})$ to be particularly convenient because it is the statistically cleanest correlator, with $\mathbf{p}_{\pi}=\mathbf{p}_D=0$.

\begin{figure*}
\includegraphics[width=.95\linewidth,clip]{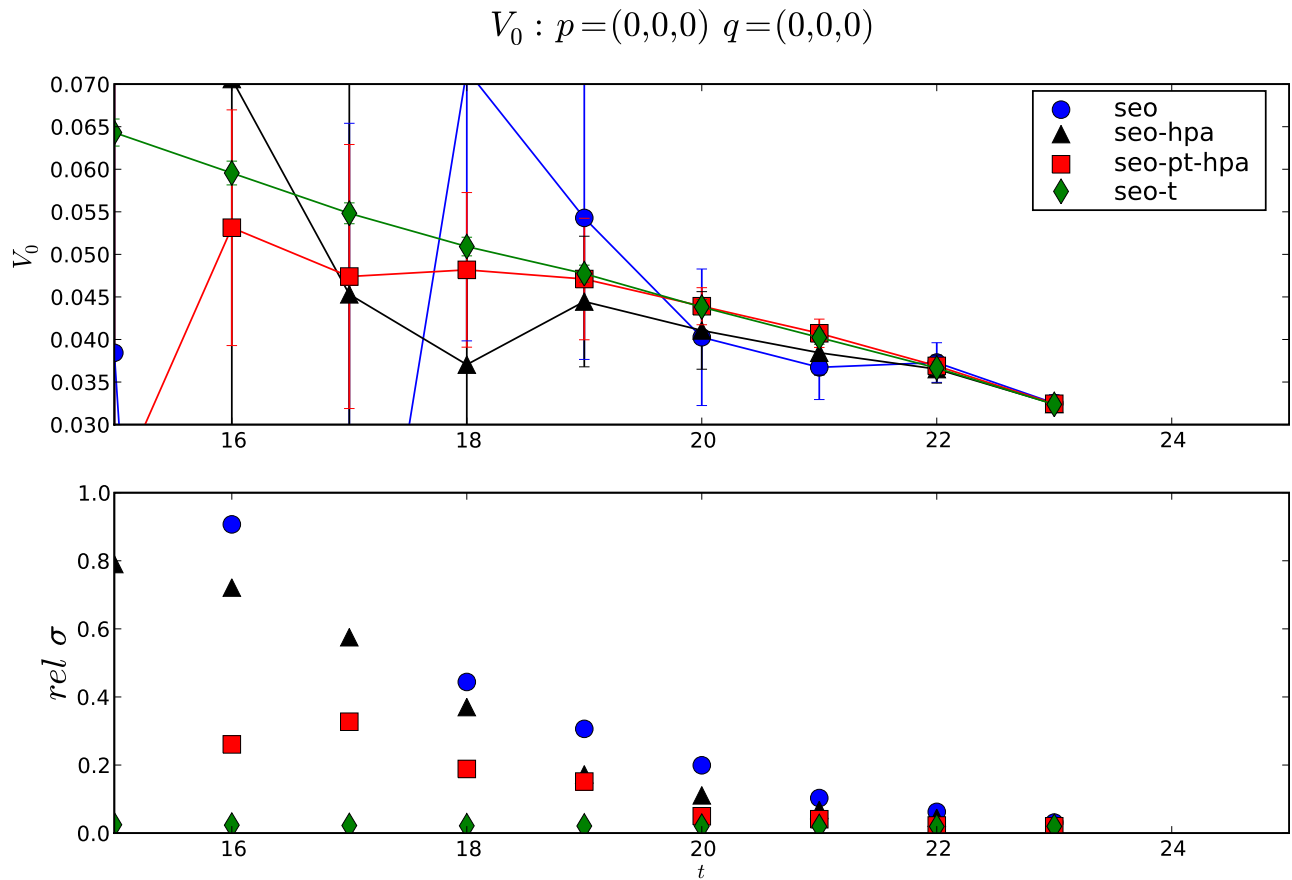}
\caption{(color online) $V_0(q^2_{\max})$ on ensemble $\mathcal L$, using three different time partitionings, in addition to spin even-odd (seo).  The pion source is at $t=0$ and the $D$ meson sink at $T=24$.  The lower plot shows the relative error of these correlators.  We show: no time partitioning (blue circles, seo), no time partitioning with HPA (black triangles, seo-hpa), partial time partitioning with HPA where the stochastic source vector has support on time slices $t=0,12,24,36$ (red squares, seo-pt-hpa), and full time partitioning (green diamonds, seo-t).  Only the pion source is smeared. }\label{tdil}
\end{figure*}

 By comparing the relative errors of $V_0(q^2_{\max})$ constructed with different combinations of time partitioning and the HPA we determined full time partitioning to be crucial for efficient noise reduction.  For this test we use spin and even-odd partitioning in addition to time partitioning, in order to reduce the number of noise vectors required to achieve a clear signal. The three different time partitionings we consider are: no time partitioning (with HPA  and without HPA), partial time partitioning with HPA (where the noise vector source has support on time-slices $T=0,12,24,36$), and full time partitioning (where the noise vector source has support on only the sink time slice $T=24$).   The results are presented in Fig.~\ref{tdil} for a single configuration from ensemble $\mathcal L$, where the $V_0(q^2_{\max})$ presented has the pion source fixed at $t=0$ and the $D$ meson sink at $T=24$ for all partitionings.

When comparing the different partitionings it is interesting to recognize that the correlators with no time and partial time partitioning can have multiple sinks $T$ at no additional cost, $0-24$ and $0,12,24,36$ respectively, whereas the full time partitioned correlator has only a single sink $T=24$.  In an attempt to acknowledge the different statistical power available in each case we construct  $V_0(q^2_{\max})$ for each partitioning from different numbers of stochastic vectors.  For no time partitioning we use 96 vectors, partial time partitioning 16 vectors, and full time partitioning 4 vectors, the simplifying assumption being that each additional time-slice provides an error reduction similar in magnitude to an additional noise vector.  

 As demonstrated in Fig.~\ref{tdil}, the noise of the correlators without full time partitioning is so much greater than the full time partitioned correlator (at fixed cost) that it is not necessary to quantify precisely the level of improvement due to having multiple sinks available: there is no way this can be competitive.  In fact, the noise increases so quickly with distance from the stochastic source that further than time-slice $t=15$ the correlators without full time partitioning lose a clear signal.  In the case of the full time partitioned correlator the noise is, however, nearly constant on all time-slices.
   Therefore we choose full time partitioning as the basis of our partitioning scheme, and note that this does not increase our cost relative to the Sequential Propagator Method, because that method is also limited to using a single sink time-slice to construct correlators. 

\begin{figure*}
\includegraphics[width=.95\linewidth,clip]{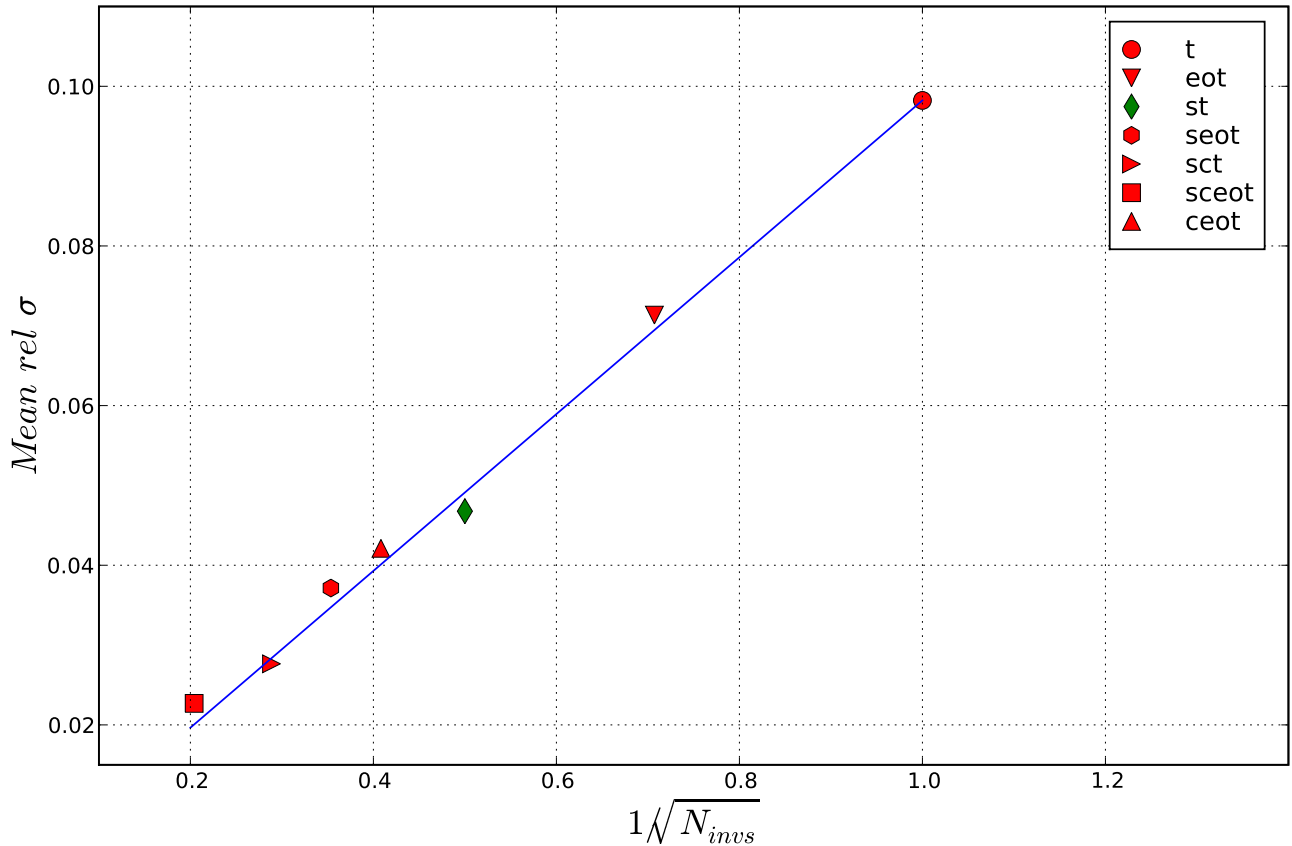}
\caption{ The relative error, averaged over the time slices, of $V_0(q^2_{\max})$ for 7 different combinations of even/odd (eo), spin (s), and color (c) partitioning as a function of the number of solves $N_{\mathrm{invs}}$ required to calculate the partitioned correlator. All correlators are fully time (t) partitioned.  These results are from a single $\mathcal L$ ensemble configuration, where only the source is smeared.}\label{dil_comp}
\end{figure*}

 Starting from full time partitioning, we then tried all combinations of spatial (even/odd), color, and spin partitioning,  as shown in Fig.~\ref{dil_comp}, again for $V_0(q^2_{\max})$ on ensemble $\mathcal L$ using a single configuration with 100 noise vectors.   The pion source is again fixed at $t=0$ and the $D$ meson sink at $T=24$ for all partitionings. The blue line represents the expected decrease of the noise with increasing the number of full time partitioned vectors.  Perhaps surprisingly, none of the alternative partitioning methods provide significant improvement over exclusively using full time dilution: spin partitioning on its own appears to have smaller errors, although the effect is small.  This result is consistent over the other configurations we examined, causing us to choose full time and spin partitioning as our preferred variance reduction technique for the rest of the paper.

\begin{figure}
\includegraphics[width=0.95\linewidth,clip]{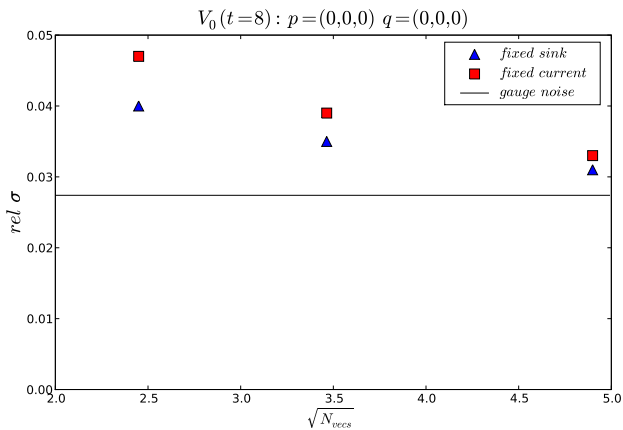}
\caption{\label{fig:err}Total error at $t=8$ for the fixed sink method, compared with the equivalent fixed current result using 6, 12, and 24 stochastic vectors. Source and sink are smeared.}
\end{figure}

 The last alteration of the calculation we examine here, which could have an effect on the noise, is placing the stochastic source at the vector current instead of at the sink.  This would result in a correlator with a fixed current and variable sink position instead of a fixed sink and variable current.  The gauge noise, and presumably the stochastic noise, is different for each approach.  For this test we considered 375 configurations from ensemble $\mathcal H$.  We present results using both approaches on a single time-slice for $V_0(q^2_{\max})$, where the $\pi$ source is at time-slice $0$, the current is at time-slice $t=8$, and the $D$ meson sink is at time-slice $T=16$.  For this particular set of source, current, and sink time-slices, which is in the middle of the plateau as shown later in the paper, the two approaches are identical up to the effects of the stochastic noise.  These results are shown in Fig.~\ref{fig:err} using 6, 12, and 24 stochastic vectors.  It can be seen that the fixed sink correlator always has smaller errors.  The gauge noise line shown is from the Sequential Propagator Method.  It is encouraging to see that even with only 6 stochastic vectors the stochastic noise is only about $\sim 30\%$ of the total noise.
 
\section{\label{comps} Comparison of methods}
\subsection{Cost estimation}
To compare the efficiency of the three methods a way of measuring the overall cost is necessary.  We choose to measure these costs $C$ in terms of heavy-quark solves required, and note that with our simulation parameters a light-quark solve $C_l$ requires approximately 30 times the computational effort of a heavy-quark solve $C_h$:
$C_l\approx 30\, C_h$
  The Sequential and Stochastic Propagator Methods will always require the same number of light-quark solves as each other, however the One-end method will require different numbers of light-quark solves depending on stochastic source location, momenta, smearings etc.~required to build the data set.  

We use $N$ fully time and spin partitioned stochastic vectors, each of which requires 4 solves to generate. The Stochastic Propagator Method generates correlators with all sink smearings, $W_{\mathrm{snk}}$, and all momenta, $p_{\pi}$ and $p_D$, with a fixed number of solves, $4N$.  It requires no additional heavy-quark solves for correlators with different spectator quark masses, $m_u$.  The Sequential Propagator Method requires 12 heavy-quark solves for every different parameter at the sink or different spectator light-quark mass.  Of course different source smearings, $W_{\mathrm{src}}$, spectator light-quark masses, $m_u$, and daughter light-quark masses, $m_l$, each require 12 additional light-quark solves to generate each point-to-all light-quark propagator.   The One-end Method requires new light-quark and heavy-quark solves for any changes in the parameters of the calculation.  

In Tab.~\ref{tab:cost} we summarize how the costs of the methods break down according to the different parameters required for a particular calculation, assuming the light-quarks are non-degenerate.  The integer factor a parameter contributes to the cost is simply labeled by the parameter in parentheses.  For example, if 2 different $\mathbf{p}_D$ were required for a calculation then $(\mathbf{p}_D)=2$.  $C_l$ and $C_h$ label the cost of a light-quark and heavy-quark solve respectively.  The Wuppertal smearings at the source (sink) can be divided over the propagator sources (sinks) in whatever manner is most efficient.  The effects of requiring the light-quarks to be degenerate are: halving the number of light-quark inversions required for the Sequential and Stochastic Propagator Methods, reducing solves required in the One-end Source approach by $4N(m_u)C_l$, and dividing the last term in the One-end Sink approach by a factor of $(m_l)$.    Note that the costs of the Sequential and Stochastic Methods only differ by the last term which is proportional to $C_h$.

\begin{table*}
\caption{ A breakdown of the costs of each method.  The integer factor a particular parameter contributes is labeled by the parameter itself, where the parameter labels are defined in the text.  Each point-to-all and sequential propagator require 12 solves, and each of the $N$ stochastic vectors require $4$ solves.  The approximate costs of the light-quark and heavy-quark solves are labeled by $C_l$ and $C_h$ respectively. $C_l\approx 30 C_h$ with our quark masses and lattice volumes.\label{tab:cost}}
\begin{ruledtabular}
\begin{tabular}{cc}      
method & computational cost \\
\hline
sequential& $12(m_l)(W_{\mathrm{src}})C_l+12(m_u)C_l +12(m_u)(\mathbf{p}_D)(W_{\mathrm{snk}})C_h$  \\
stochastic& $12(m_l)(W_{\mathrm{src}})C_l+12(m_u)C_l+ 4NC_h$ \\
one-end (source)& $4N(\mathbf{p}_{\pi})(m_l)(W_{\mathrm{src}})C_l+4N(m_u)C_l+  48N(m_u)(\mathbf{p}_D)(W_{\mathrm{snk}}) C_h $\\
one-end (sink)& $4N(\mathbf{p}_{D})(W_{\mathrm{snk}})C_h+4N(m_u)C_l+ 48N(m_u)(\mathbf{p}_{\pi})(m_l)(W_{\mathrm{src}})C_l $\\
\end{tabular}
\end{ruledtabular}
\end{table*}

\subsection{One-end Method}
First we compare correlators generated by the One-end Sink Method to correlators generated from the Sequential and Stochastic Propagator Methods.    For the One-end Method we use only a single, full time and spin partitioned, stochastic vector per configuration.  Only a single stochastic vector is necessary because the properties of Eq.~(\ref{eq:z2}) are valid across different configurations, as well as within a single configuration.  Of course more stochastic vectors could be used, but as Tab.~\ref{tab:cost} shows the cost grows very rapidly with additional stochastic samples.  For the relatively cheap heavy-quark stochastic propagator required for the Stochastic Propagator Method we use $N=24$ fully time and spin partitioned vectors.  The exact costs of the methods are entirely dependent on the needed final data set, but here we demonstrate that the One-end Method is, in any ``realistic" calculation, far less efficient than the other 2 methods.    

In general we expect the stochastic error of the One-end Sink and Source Methods to be different, and in fact found the One-end Sink Method to have smaller errors.   According to Tab.~\ref{tab:cost} the Sink procedure is likely to be much more expensive than the Source procedure, but we defer discussion of the cost until the end of this subsection.  

 A comparison of the three methods on 250 configurations can be seen in Figs.~\ref{x3a}--\ref{x3c} on ensemble $\mathcal L$, where all correlators are smeared at the source and sink. Fig.~\ref{x3a} shows $V_0(q^2_{\max})$ for the three methods.  In this figure it can be seen that the Stochastic Propagator and Sequential Propagator $V_0(q^2_{\max})$ have smaller errors than the One-end $V_0(q^2_{\max})$.  In Fig.~\ref{x3b}, a plot of the $V_1(|\mathbf{p}|=0,|\mathbf{q}|=1)$ correlator where a single momentum is used, shows the correlator for the One-end method to have the smallest errors.  This might be expected due to the explicit momentum projection at the source.
This is too naive a comparison though, because the other two methods have 5 more rotationally equivalent correlators available at negligible cost, which can be averaged. Fig.~\ref{x3c} shows the effect of averaging the 6 correlators available when using the Stochastic and Sequential Methods.  The One-end generated data, which have no rotationally equivalent correlators to average over, now have slightly larger errors than the averaged data from the other two methods.   

  We have performed similar comparisons for all other momentum combinations we are interested in and also for selected momentum combinations on ensemble ${\mathcal H}$, finding similar results in all cases.  Considering there is no reduction in errors using the One-end Method and its high cost in a realistic calculation, it is obvious that the One-end Method is far less efficient than the other methods.  For example, consider a calculation with degenerate light-quarks where $(m_u)=(m_l)=3$, $(\mathbf{p}_D)=(\mathbf{p}_{\pi})=3$, $(W_{\mathrm{src}})=(W_{\mathrm{snk}})=1$, and $C_l = 30C_h$:
 \begin{widetext}
 \begin{center}
  \begin{tabular}{lll}
Cost(sequential) & $=12(3)(30\,C_h)+12(3)(3)(C_h)$ & $ = 1188\,C_h  $ \\
Cost(stochastic) & $=12(3)(30\,C_h)+96(C_h)$ & $= 1176\,C_h$ \\
Cost(one-end source) & $= 4(3)(3)(30\,C_h)+4(3)(30\,C_h)+48(3)(3)(C_h)$ & $= 1872\,C_h$ \\
Cost(one-end sink) & $= 4(3)(C_h)+4(3)(30\,C_h)+48(3)(3)(30\,C_h)$ & $= 133372\,C_h.$ \\
  \end{tabular}
  \end{center} 
  \end{widetext}
  As can be seen in the calculation above we chose a very expensive One-end procedure to perform this calculation, the One-end Sink Method; however, even the One-end Source Method would be much costlier than the Sequential or Stochastic Propagator Method.  We mention again that the One-end Sink Method  was chosen to minimize the error, and so the cheaper One-end Source Method would be even noisier.  The One-end Methods at these quark masses and volumes are evidently not close to being competitive with the other two approaches, and we drop these methods from further consideration.

\begin{figure}
\subfloat[(color online) The upper plot shows the scaled temporal component of the vector matrix element, $V_0$, at $q^2=q^2_{\max}$, from the three different methods using  250 configurations.  The lower plot shows the relative error of the three correlators.  The Stochastic Propagator based three-point function is labeled by $spin-z_2$ (blue circles), the Sequential Propagator by $seq$ (red squares), and the one-end by $one-end$ (green triangles).]{\includegraphics[width=\linewidth,clip]{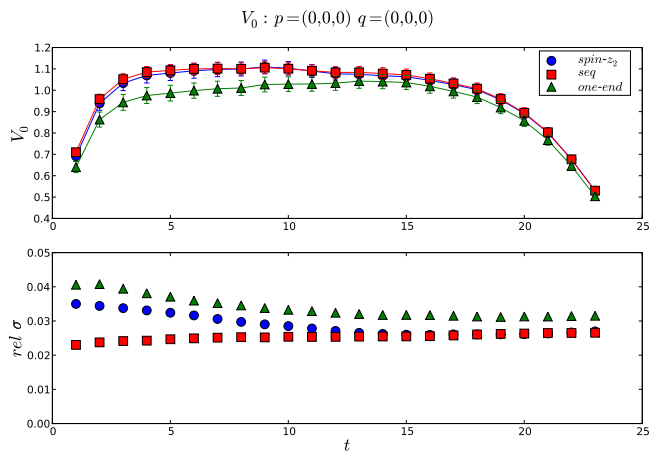}\label{x3a}}
\caption{Error comparison of the three methods.  250 configurations from Ensemble ${\mathcal L}$ are used.\label{figx3}}
\end{figure}

\begin{figure}
\ContinuedFloat
\subfloat[ $V_1$]{\includegraphics[width=\linewidth,clip]{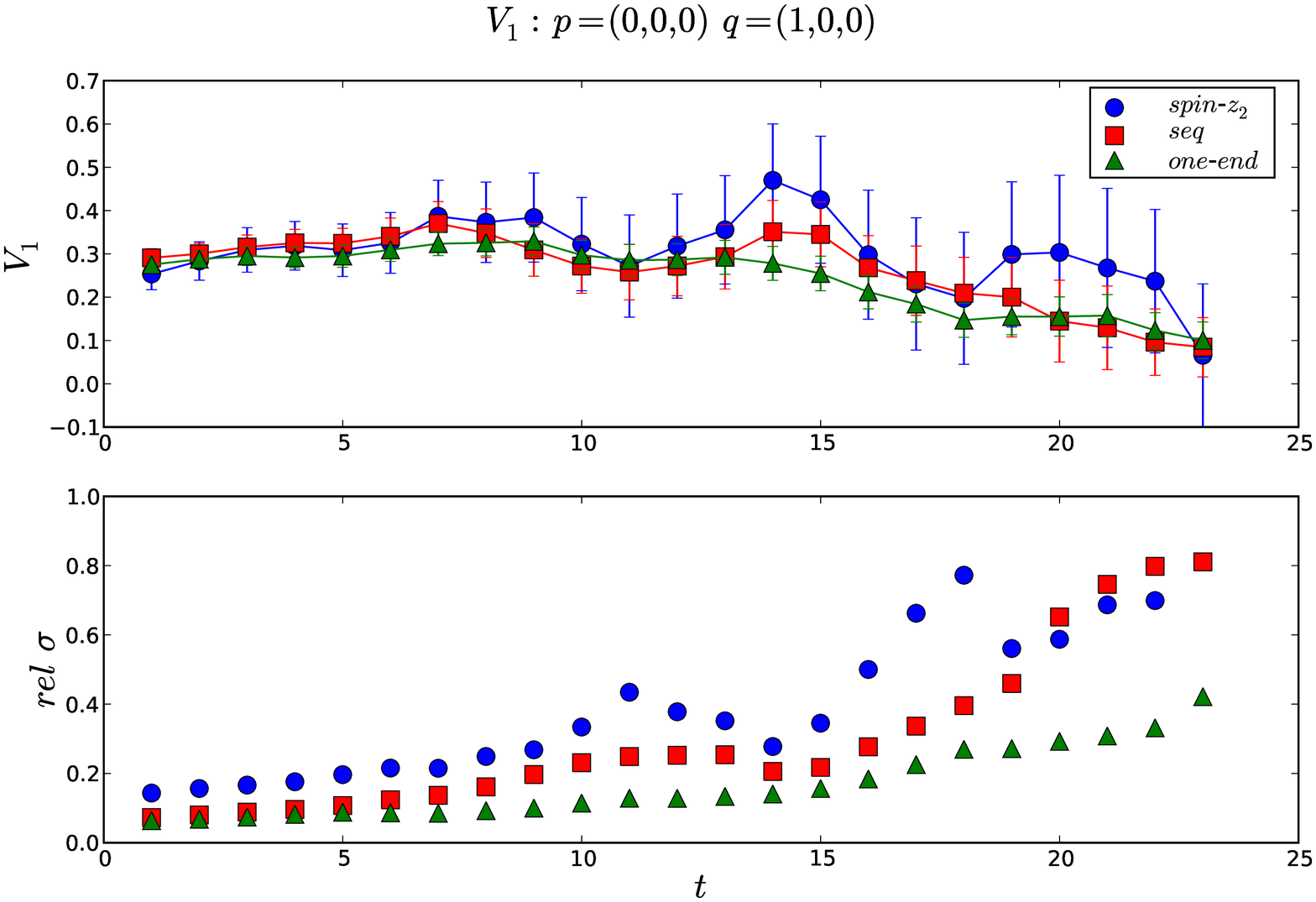}\label{x3b}}\\
\subfloat[average over $V_i,i=1,2,3$]{\includegraphics[width=\linewidth,clip]{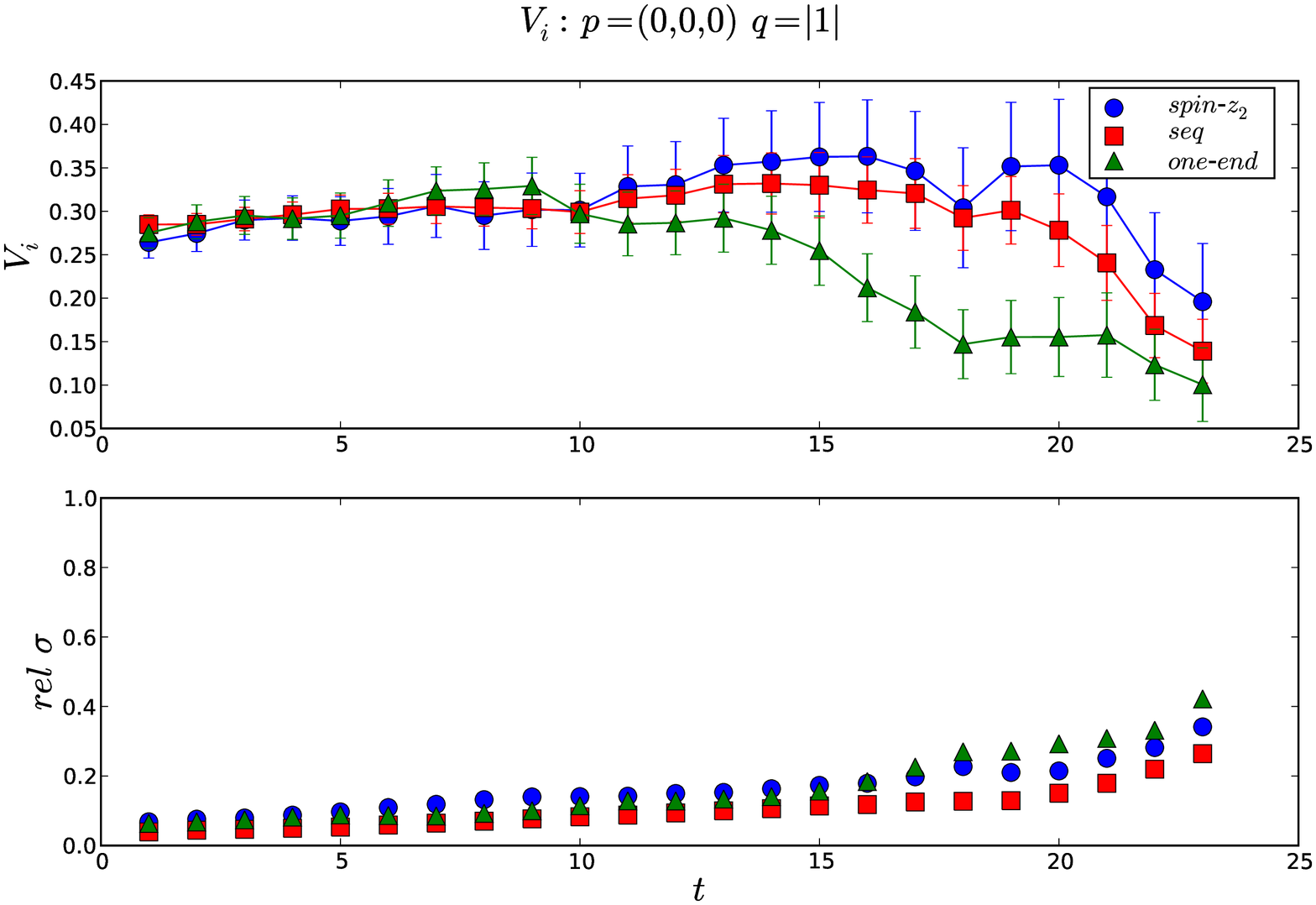}\label{x3c}}
\caption{cont. Error comparison of the three methods.  250 configurations from Ensemble ${\mathcal L}$ are used.\label{x3}}
\end{figure}

\subsection{Sequential Propagator versus Stochastic Propagator}
In this section we compare the two more promising methods, as suggested by the plots and rough cost calculation from the last subsection.   We perform a direct comparison of the Sequential and Stochastic Propagator Methods to determine when it is advantageous to use the Stochastic Propagator Method.  

\begin{figure}
\subfloat[(color online) $V_0(q^2_{\max})$ from 100 configurations on ensemble $\mathcal H$.  The Stochastic Method (purple circles) and Sequential Method (green squares) correlators on the left are not smeared.  The Stochastic Method (blue circles) and Sequential Method (red squares) correlators on the right are smeared at source and sink. ]{\includegraphics[width=.98\linewidth,clip]{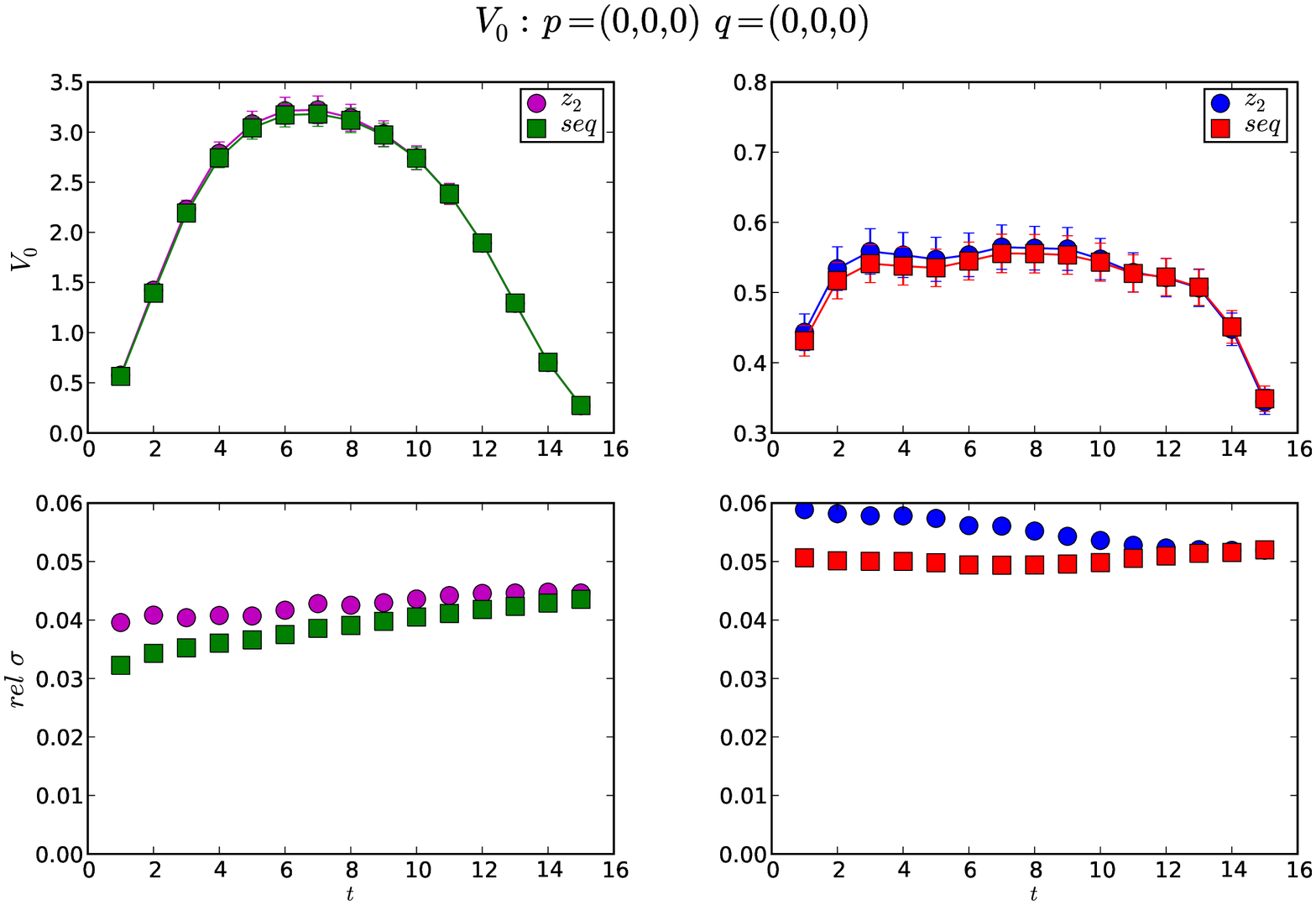}
\label{smrcomp:a}}\\
\subfloat[(color online) $V_0(q^2_{\max})$ from 220 configurations on ensemble $\mathcal L$.  The Stochastic Method (purple circles) and Sequential Method (green squares) correlators on the left are smeared at the source but not the sink.  The Stochastic Method (blue circles) and Sequential Method (red squares) correlators on the right are smeared at source and sink. ]{\includegraphics[width=.98\linewidth,clip]{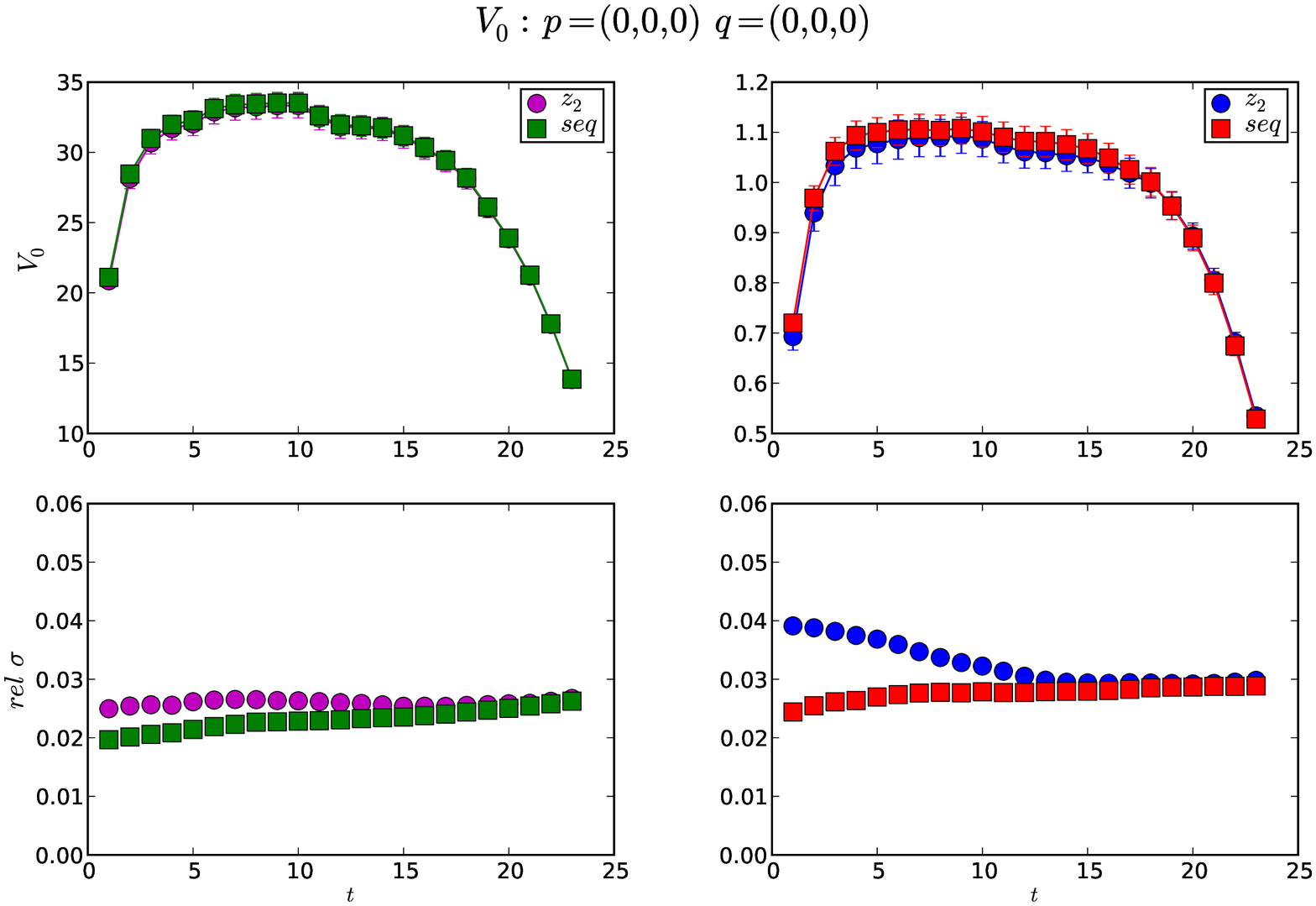}
\label{smrcomp:b}}
\caption{\label{smrcomp} We compare the effects of smearing on the $V_0(q^2_{\max})$ correlator on the two ensembles.  The smearing dependence of the wave function amplitudes, $Z_P$ and $Z_H$ in Eq.~(\protect\ref{v0}), causes the difference in scale for the different smearings. }
\end{figure}

As a first step in an effort to understand the behavior of the noise present in both methods we examine the effects of smearing. $V_0(q^2_{\max})$ with two different smearings is shown in Figs.~\ref{smrcomp:a}--\ref{smrcomp:b} for the two ensembles.  We remind the reader that in this setting, in the limit $N\to \infty$, the Stochastic Method will yield exactly the same result as the Sequential Method.  The results from 100 configurations on ensemble $\mathcal H$ are presented in Fig.~\ref{smrcomp:a}, and for 220 configurations on ensemble ${\mathcal L}$  in Fig.~\ref{smrcomp:b}.  The behavior of the gauge noise can be directly seen from the Sequential Propagator Method correlators.  It decreases slightly with distance from the $D$ meson sink regardless of smearing.   The contribution of the stochastic error can be determined by comparing with this purely gauge error: for all smearings and both ensembles it is negligible compared to the gauge noise near the stochastic source at the $D$ meson sink, but grows mildly with distance from the source.  The effects of smearing on the gauge and stochastic noises are both mild, and smearing clearly helps to produce cleaner plateaus.  

We now compare the total errors of both methods using the five statistically cleanest combinations of correlators with finite momenta on ensemble $\mathcal L$, where each combination of correlators corresponds to a particular $f_0(q^2)$ and $f_+(q^2)$.  When averaging over so many equivalent correlators it is more straightforward to extract the form factors from the appropriate combination of matrix elements and kinematic factors, and then perform the average.  The energies needed for this construction are determined from bootstrap ensembles of two-point fits. These are then combined with the corresponding bootstrap ensembles of the three-point correlators.  

We assume only three $D$ meson momenta are desired  (on these lattices, $D$ mesons with momentum $|\mathbf{p}_D|>\sqrt{2}$ have significantly distorted dispersion relations): 
\begin{equation}
\mathbf{p}_D=(0,0,0),(1,0,0),(1,1,0)\,.
\end{equation}
We average over all rotationally equivalent correlators to improve the statistics; however, 
there are many correlators with rotationally equivalent sink momenta which are not available to the Sequential Propagator Method but are essentially free for the Stochastic Propagator Method.  These additional momenta are utilized in the stochastic case.   We also consider the same momentum range for the pion (on these lattices and valence quark masses correlators with $|\mathbf{p}_{\pi}|>\sqrt{2}$ are too noisy to produce a clear signal).  Both methods have the same rotationally equivalent source momenta available.  
Figs.~\ref{fig:a}--\ref{fig:e} are representative of the range of behavior of the relative errors in our data set. The number of correlators averaged over is shown below each figure for both methods.

 The Sequential Propagator Method uses $12 \times 3=36$ heavy-quark inversions to generate the correlators.  The stochastic propagator method  uses $(N=24) \times 4=96$. As suggested in Figs.~\ref{fig:err} and \ref{smrcomp:a}--\ref{smrcomp:b}, less vectors could be used without a significant increase of the error. However, as seen in the cost estimation in the previous section of a realistic calculation, choosing $N=24$ is still cheaper (although modestly) than the Sequential Method.  If we performed a naive fixed cost comparison with no assumptions about what constitutes a realistic data set, only 9 spin diluted vectors (requiring 36 heavy-quark solves) could be used, and the stochastic contribution to the total error in the Stochastic Propagator Method would increase approximately by a factor of 1.6.  As the stochastic error is a small percentage of the total error (the worst case appears in Fig.~\ref{smrcomp:b} in time slices 1-10, where the stochastic error accounts for $\sim 5\%$ of the total error) the Stochastic Propagator Method would still be competitive with the Sequential Propagator Method for most data points.  
  
\begin{figure}
\subfloat[Stoch - 6, Seq -1 : $\mathbf{p}=(1,0,0),\mathbf{q}=(1,0,0)$] {\label{fig:a}\includegraphics[width=\linewidth,clip]{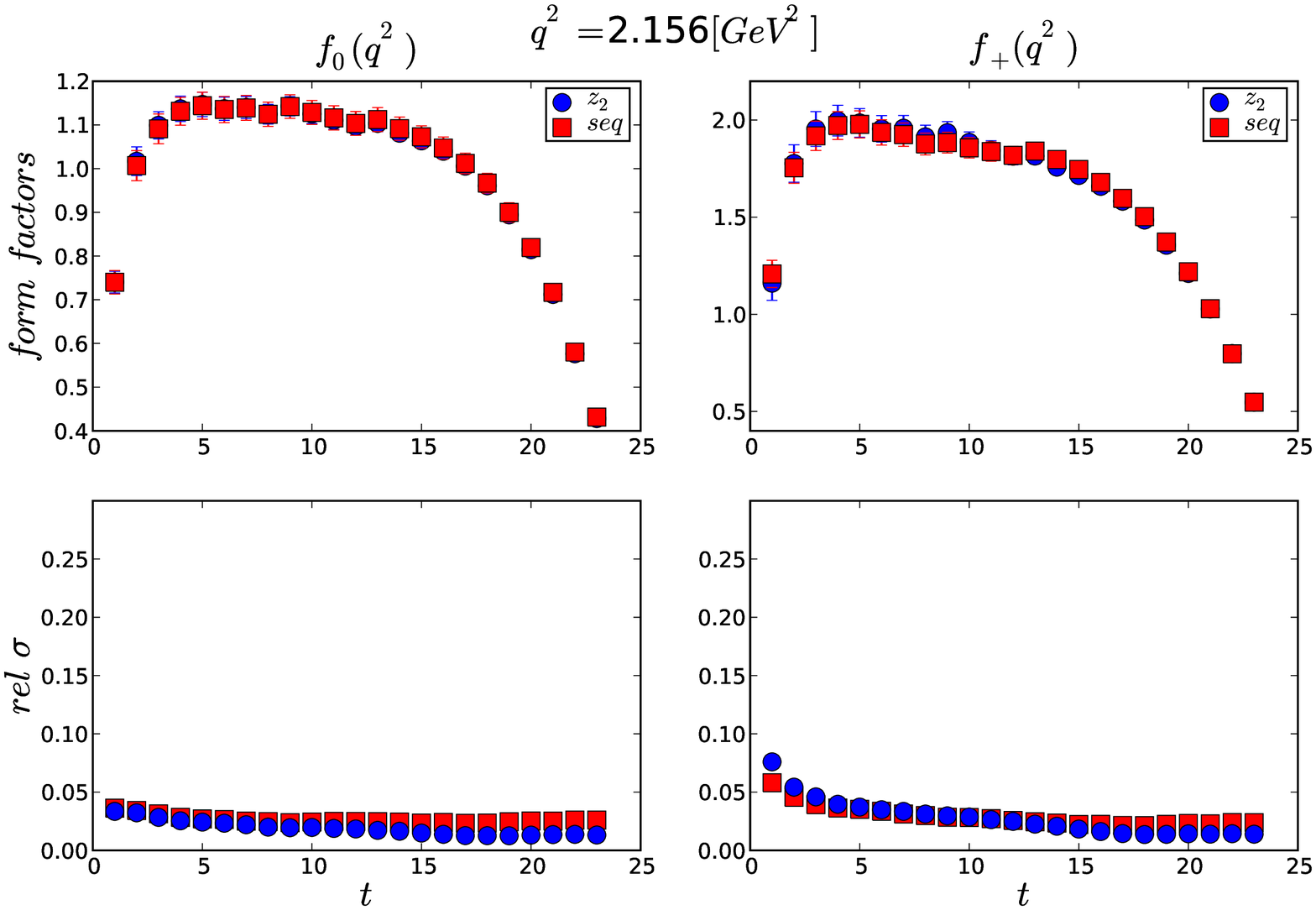}}\\
\subfloat[Stochastic - 12, Sequential - 1 : $\mathbf{p}=(1,1,0),\mathbf{q}=(1,1,0)$] {\label{fig:b}\includegraphics[width=\linewidth,clip]{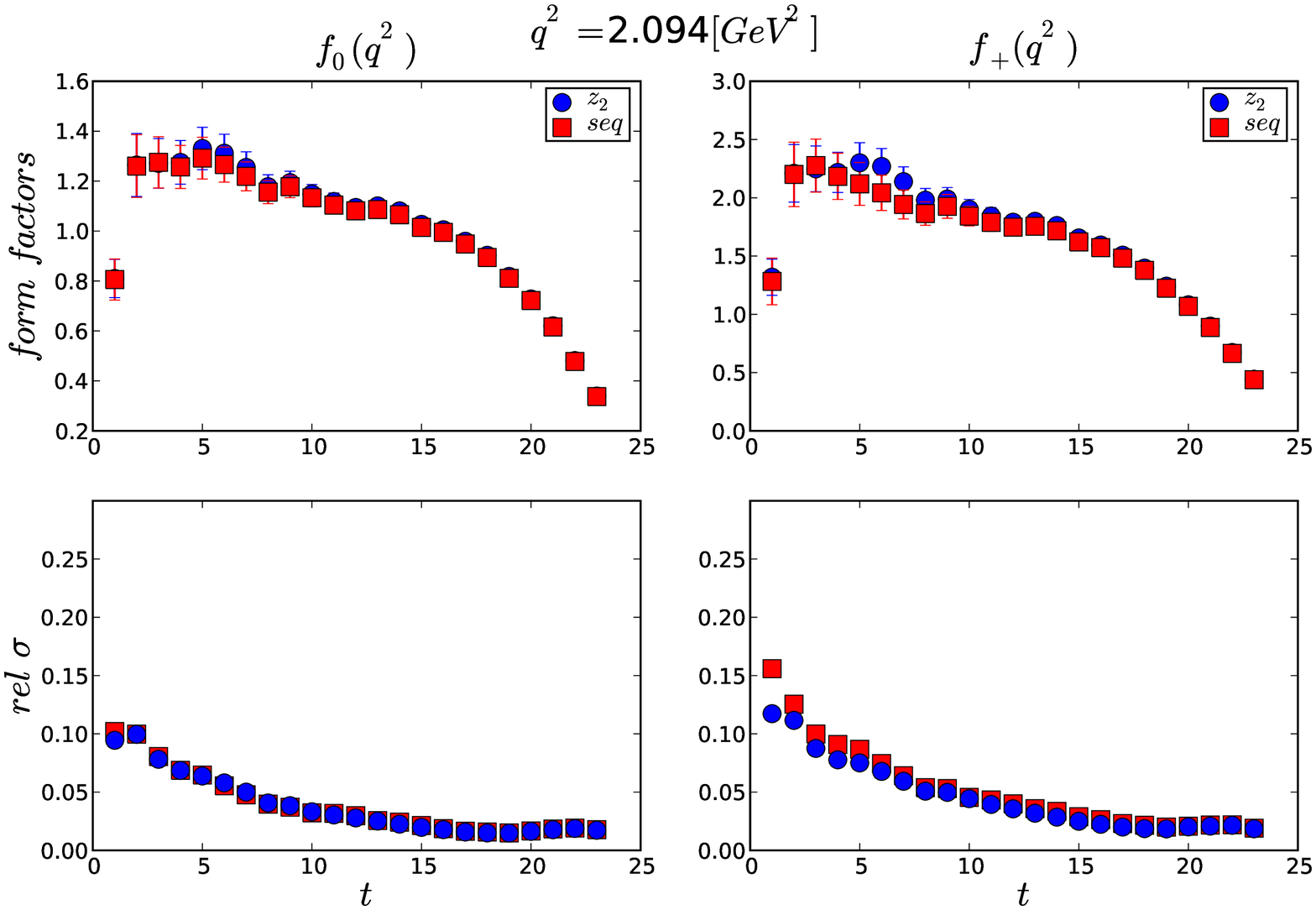}}
\caption{Comparison of errors from correlators generated using the Stochastic and Sequential Methods.  The data in the above plots are generated using 668 configurations.  The spatial momentum corresponding to the $q^2$ of the data is shown in each sub-caption.  The number of rotationally equivalent correlators available (and used) for averaging is listed after each method's name in the subcaptions. \label{fig:ae}}
\end{figure}

\begin{figure}
\ContinuedFloat
\subfloat[Stochastic - 6, Sequential -1  $\mathbf{p}=(1,0,0),\mathbf{q}=(0,0,0)$] {\label{fig:c}\includegraphics[width=\linewidth,clip]{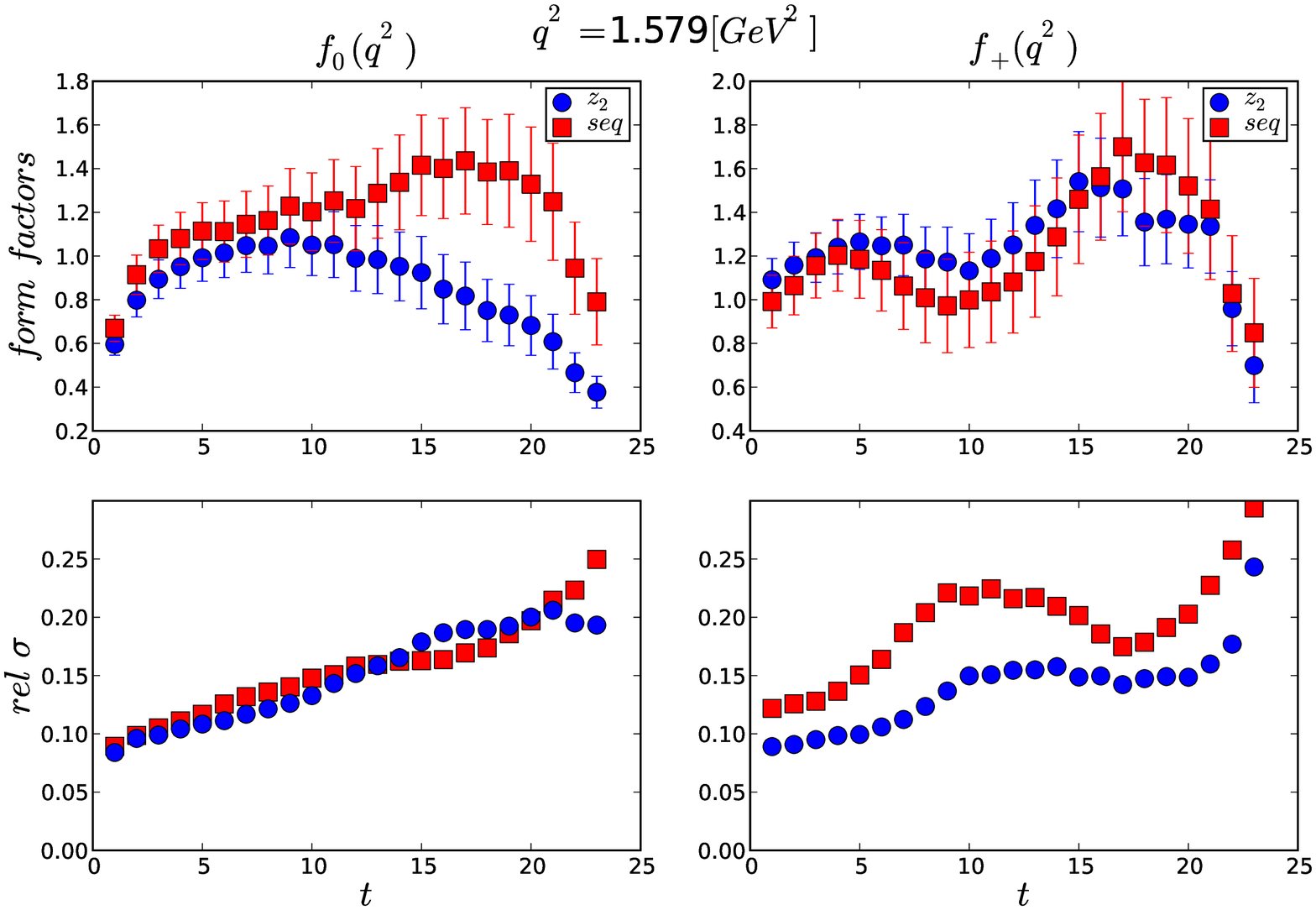}}\\
\subfloat[Stochastic - 24, Sequential - 2 $\mathbf{p}=(1,1,0),\mathbf{q}=(1,0,0)$] {\label{fig:d}\includegraphics[width=\linewidth,clip]{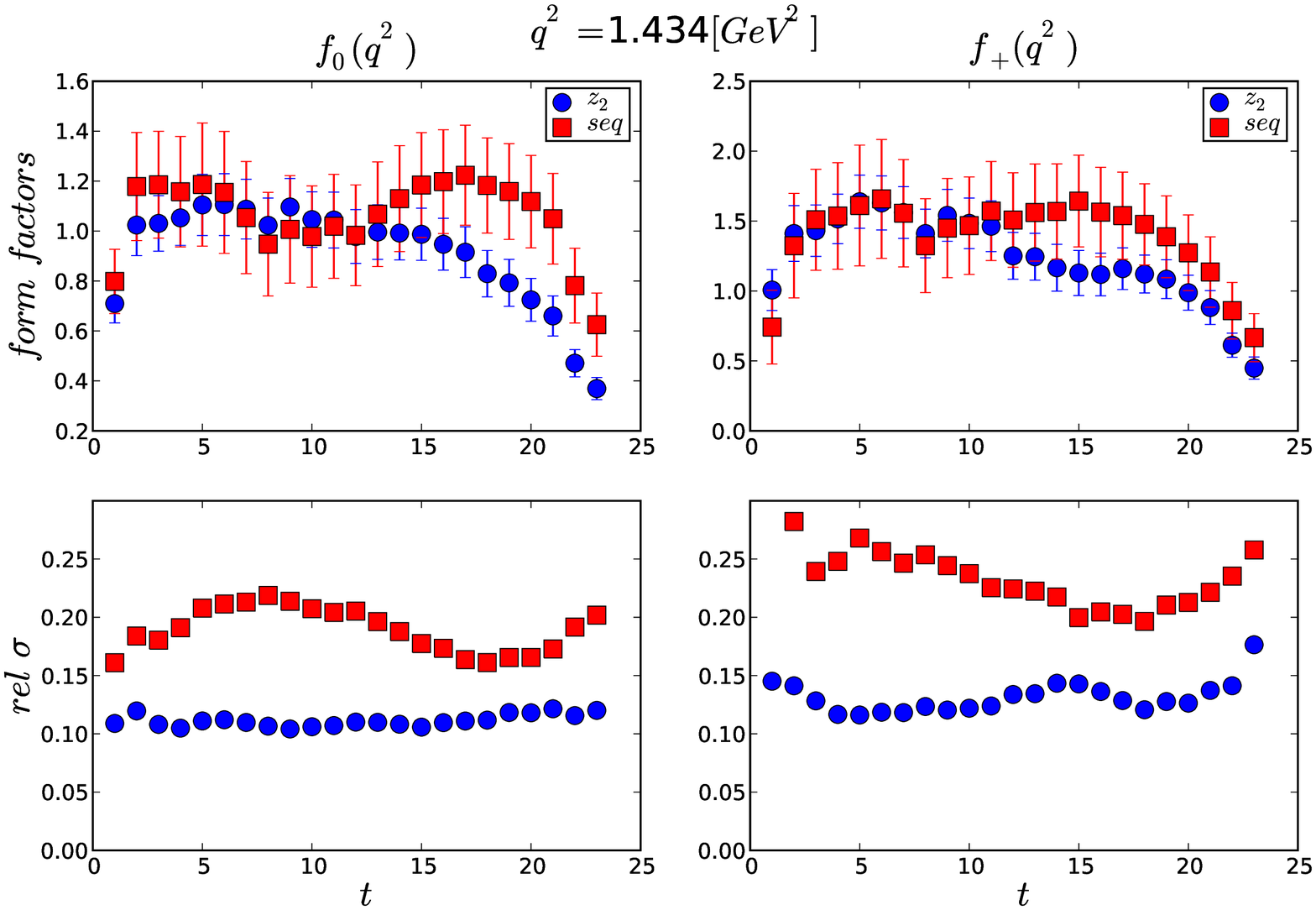}}\\
\caption{cont. Comparison of errors from correlators generated using the Stochastic and Sequential Methods.\label{fig:aee}}
\end{figure}

\begin{figure}
\ContinuedFloat
\subfloat[Stochastic - 6, Sequential - 6 $\mathbf{p}=(0,0,0),\mathbf{q}=(1,0,0)$] {\label{fig:e}\includegraphics[width=\linewidth,clip]{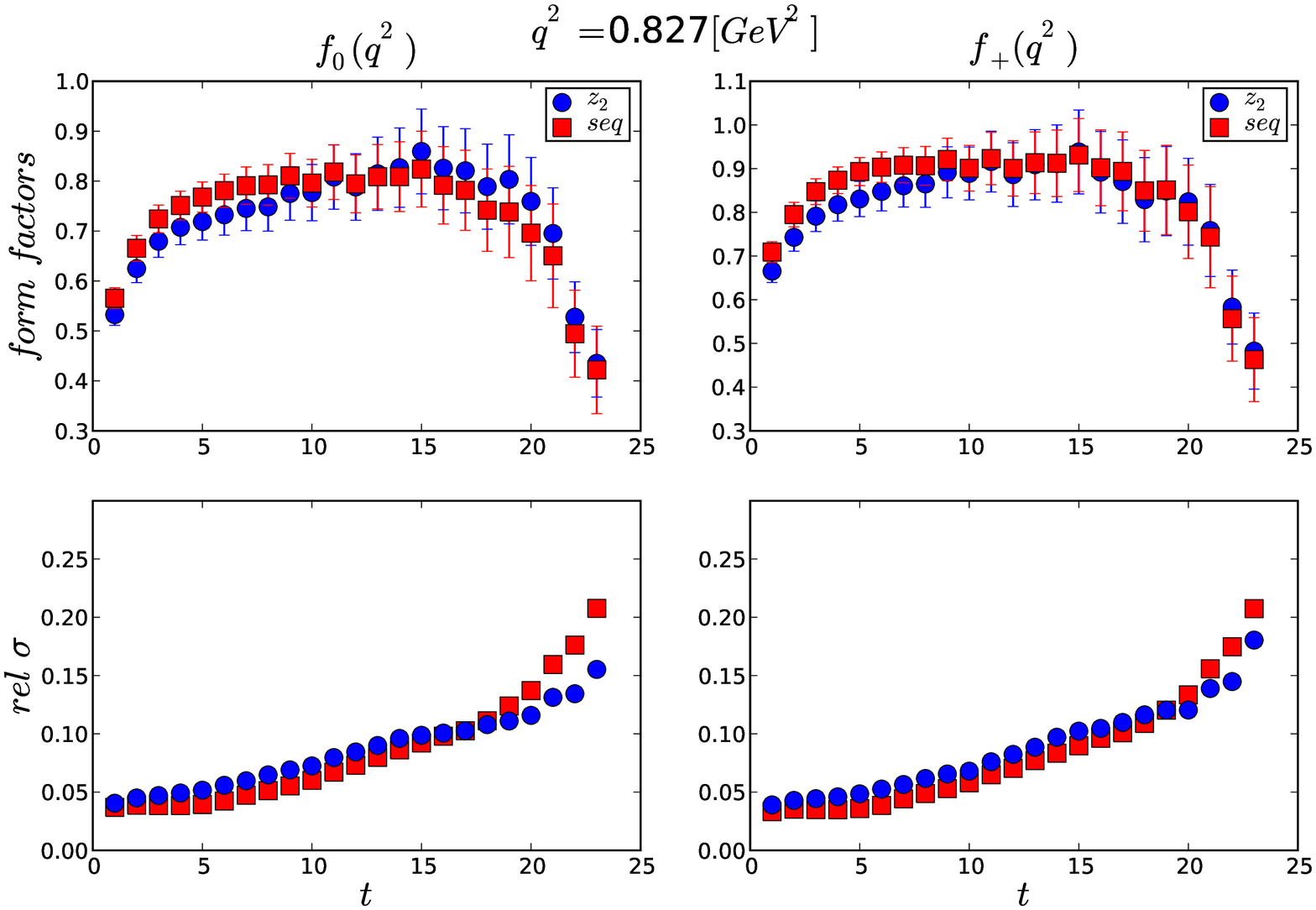}}
\caption{cont. Comparison of errors from correlators generated using the Stochastic and Sequential Methods. \label{fig:aeee}   }
\end{figure}

We have examined all correlators with an appreciable signal and draw the conclusion that with the modest number of 24 full time and spin partitioned stochastic estimates the error in the stochastic correlators is at worst comparable to the error in the sequential correlators, while at best the additional data available result in smaller total errors for the Stochastic Propagator Method.  It is also clear that the Stochastic Propagator Method is much more flexible, because additional sink smearings, sink interpolating fields, and spectator quark masses do not add significantly to the cost of the calculation.   In the next section we use the Stochastic Propagator Method generated correlators to perform a $q^2\to 0$ extrapolation.

\section{Matching and results}
In order to connect our results to observables of phenomenological interest we perform the matching and ${\mathcal O}(a)$ improvement of the vector current.  The matching calculation takes the form, 
\begin{equation}
V_{\mu}^{\mathrm{cont}}(q^2)=Z_V\left[V_{\mu}(q^2)+aic_V\partial_{\nu} T_{\mu\nu}(q^2)\right]\,,
\end{equation}
where $T_{\mu\nu}=\bar{\psi}_c\sigma_{\mu\nu}\psi_l$ is the tensor current  with $\sigma_{\mu\nu}=\frac i 2 [\gamma_{\mu},\gamma_{\nu}]$.  The matching factor $Z_V$ is known non-perturbatively \cite{Bakeyev:2003ff, Gockeler:2010yr}, while the coefficient of the improvement term, $c_V$, is known to one-loop in perturbation theory \cite{Sint:1997jx}. We have examined the effects of ${\mathcal O}(a)$ improvement and they do not change the behavior of the noise.  

It is interesting to note that a similar relation to that used in Ref.~\cite{Na:2009au}, where $f_0(q^2)$ can be directly related to the scalar current, holds to the order at which we are working.  The observation the authors make in Ref.~\cite{Na:2009au} is that the continuum Ward Identity,
\begin{equation}
q^{\mu}\langle \pi | V^{\mathrm{cont}}_{\mu}| D \rangle=(m_c-m_l)\langle \pi | S^{\mathrm{cont}}| D_l\rangle\,,
\end{equation}
is valid on the lattice up to a multiplicative matching factor.  For the HISQ action \cite{Follana:2006rc}, which they use, they obtain,
\begin{equation}
Z_V q^{\mu}\langle \pi | V^{\mathrm{lat}}_{\mu}| D \rangle=(m_c-m_l)\langle \pi | S^{\mathrm{lat}}| D_l\rangle\,.
\end{equation}
This relation holds because the right hand side is renormalization group invariant when the same action is used for both valence quarks. 

 Due to the anti-symmetry of the tensor operator this relationship is also valid when using Wilson-type quarks, and allows us to calculate $f_0(q^2)$:
\begin{equation}       
f_0(q^2)=\frac {m_c-m_l}{m_D^2-m_{\pi}^2}\langle S^{\mathrm{lat}} \rangle +{\mathcal O}(a^2)\,,
\label{f0tos}
\end{equation}
 where $f_0(0)=f_+(0)$.  We extract $f_0(0)$ using this approach as well as using the conventional approach with the vector current.

 As in the comparisons shown in Fig.~\ref{fig:ae}, to extract the form factors we first combine the appropriate three-point correlators, which are proportional to the vector matrix element,  using the same bootstrap samples for all rotationally equivalent components, so that the resulting correlators in time are proportional to the form factors:
\begin{equation}
C_3(T,t;\mathbf{p}_H,\mathbf{q})\stackrel{T\gg t\gg 0}{\longrightarrow} \frac {Z_P}{2E_P} \frac{ Z_H}{2E_H} f_{0,+}  e^{-E_P t} e^{-E_H(T-t)}\,.
\end{equation}
We then form the ratio from the bootstrap ensembles of our two-point and three-point functions,
\begin{equation}
f_{0,+}=Z_{\pi}Z_{D}\frac {C_3(T,t;\mathbf{p}_H,\mathbf{q})} {C_{\pi}(t) C_{D}(T-t)}\,,
\end{equation}
and remove $Z_{\pi}$ and $Z_D$ using our two-point fit results.  It can be seen in Fig.~\ref{fig:ae} that the plateaus in the first two correlators are not clear.  The lack of clear plateaus requires us to include excited states in our fits.  To do this we determine the energy differences between the first excited state and ground state from two-state fits to the two-point functions.  Using the energy differences we can perform a three parameter fit to extract the form factors,
  \begin{equation}
  p_0+p_1e^{-\Delta E_{\pi}t}+p_2 e^{-\Delta E_{D}(T-t)}\,,
  \label{plat}
  \end{equation}   
and obtain reasonable $\chi^2$ values.  Although plateau fits were suitable for the remaining correlators we found additional time-slices and wider fitting windows could be included using Eq.~(\ref{plat}) for all correlators, and use this fit model for all form factor results presented here. 

\begin{figure}
\includegraphics[width=.95\linewidth,clip]{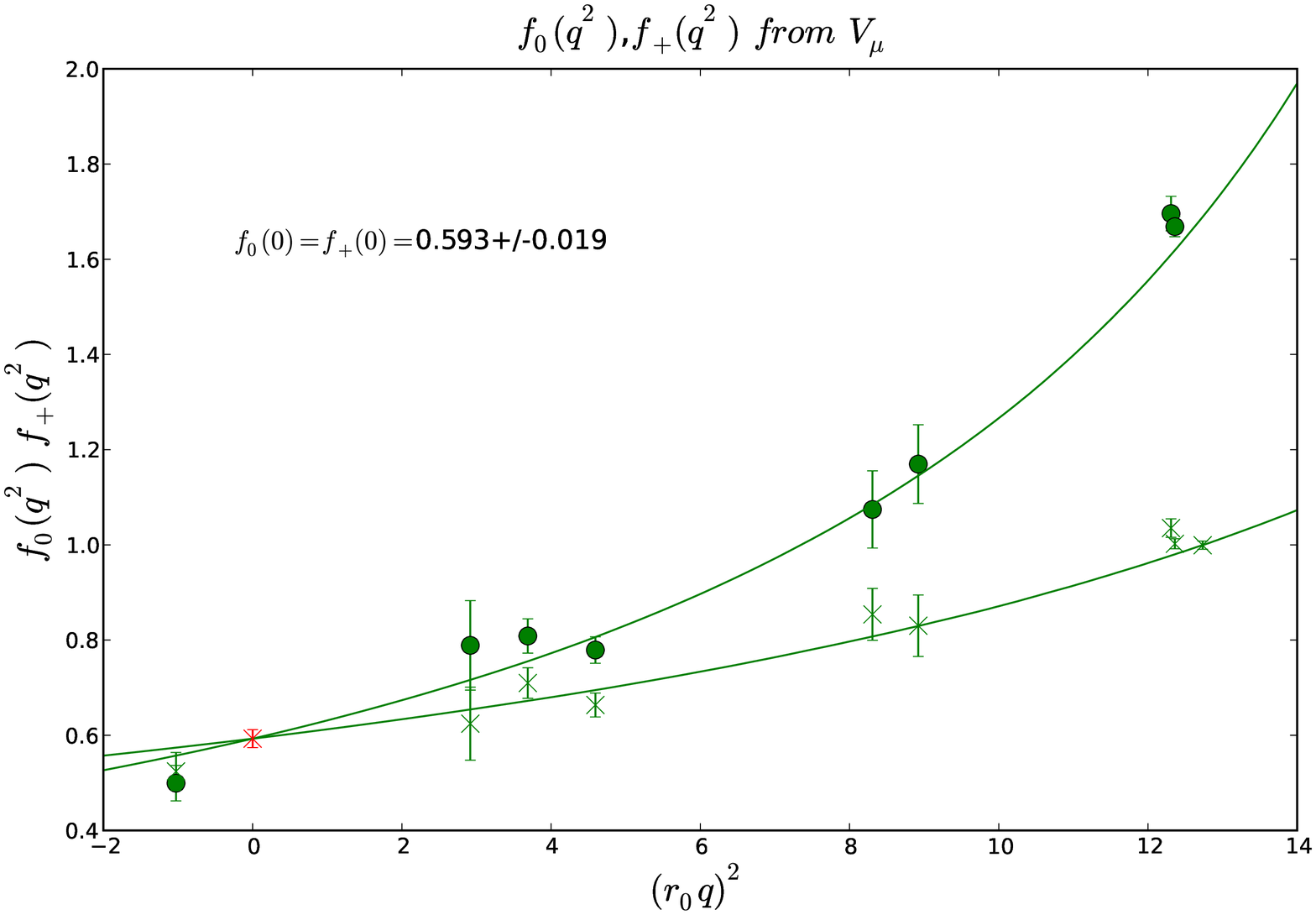}
\caption{Results extracted from $\langle V_{\mu} \rangle$,
using the Stochastic Propagator Method on the $24^3\times 48$ ensemble with 24 spin diluted stochastic vectors on 539 configurations $\times$ 4 time sources.} \label{ff24}
\end{figure}

\begin{figure}
\includegraphics[width=.95\linewidth,clip]{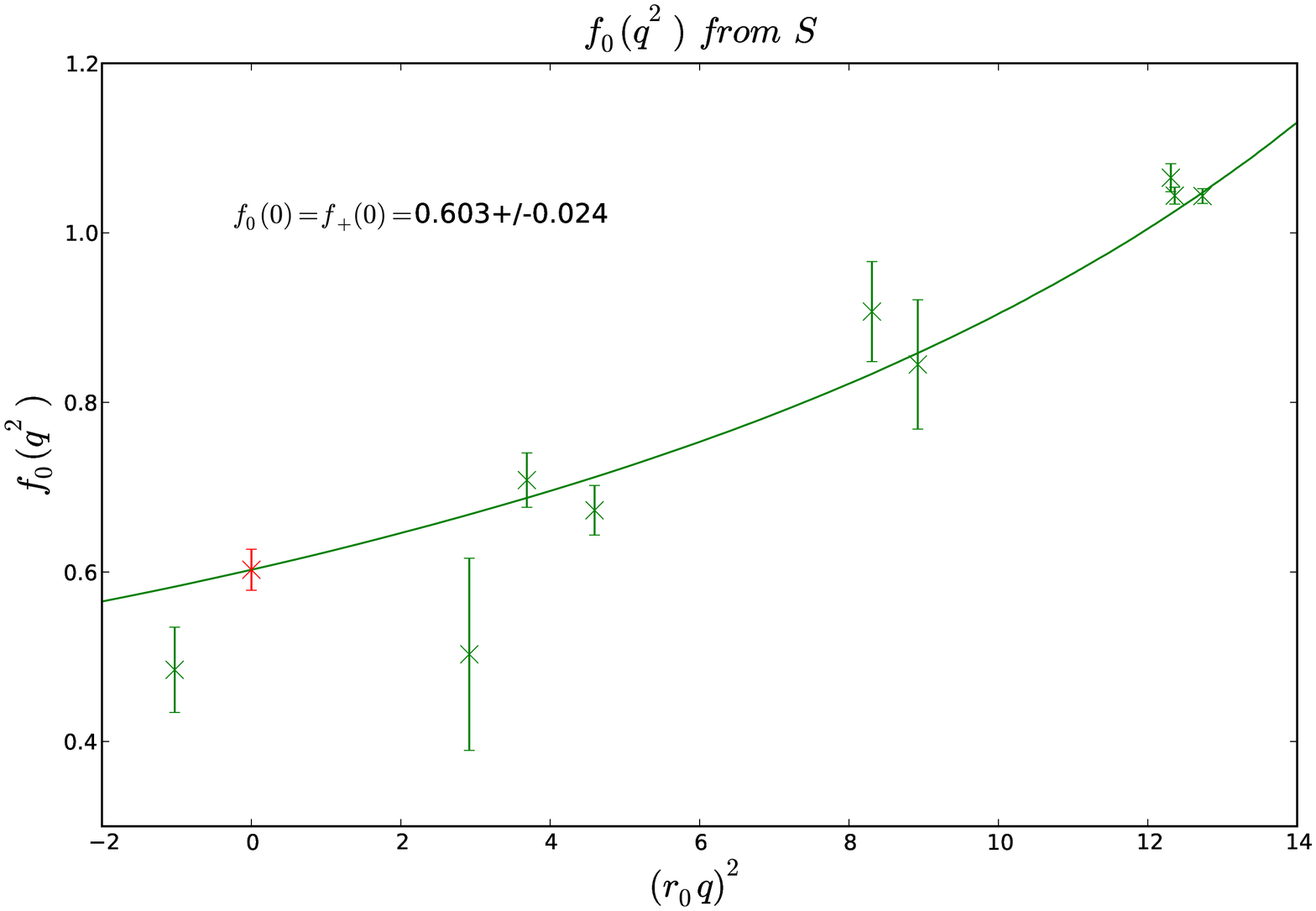}
\caption{Results extracted from $\langle S \rangle$, using the Stochastic Propagator Method on the $24^3\times 48$ ensemble with 24 spin diluted stochastic vectors on 539 configurations $\times$ 4 time sources.} \label{ffs24}
\end{figure}

Renormalized and $\mathcal{O}(a)$ improved lattice results using
the Stochastic Sink Method for $f_+(q^2)$ and $f_0(q^2)$ calculated
on the $24^3\times 48$ ensemble are presented in Fig.~\ref{ff24}.  The form
factors are plotted versus $(r_0 q)^2$, where
$r_0^{-1} \approx 0.422$ GeV \cite{Gockeler:2008we}.  530 configurations with
24 stochastic vectors and 4 different time sources on every lattice were used.
We also present results, in Fig.~\ref{ffs24}, for the same calculation
using the relation Eq.~(\ref{f0tos}).  This is advantageous because
no matching coefficients are required,
removing a source of systematic
uncertainty.   The two approaches to constructing the $f_0$
correlators are in all cases comparable within statistical errors;
however, we find the statistical errors of the scalar current based
data to be slightly larger.
The $q^2\to 0$ extrapolation is performed
using the Be\'{c}irevi\'{c} and Kaidalov
parameterization \cite{Becirevic:1999kt},
\begin{equation}
f_0(q^2)=\frac {c(1-\alpha)} {1-\tilde{q}^2/\beta}\,, \qquad f_+(q^2)=\frac{c(1-\alpha)}{(1-\tilde{q}^2)(1-\alpha \tilde{q}^2)}\,,
\end{equation}  
with $\tilde{q}=q/m_D^*$. $m_D^*$ is the mass of the heavy-light vector meson and $c$, $\beta$, $\alpha$ are fit parameters.  Reasonable $\chi^2$ were obtained using $\langle V_{\mu} \rangle$ or $\langle S \rangle$, with $f_+(0)=0.593(19)$ and $f_+(0)=0.603(24)$, respectively.  The errors are statistical only.
Of course to extract physical results from any of these methods we would need to perform a chiral and lattice spacing extrapolation. We note that these results are in agreement with the value  $f_+(0)=0.64(3)(6)$, obtained in  Ref.~\cite{Aubin:2004ej}, the most recent lattice calculation where systematic errors are estimated. 

\section{Conclusions}
We have studied two stochastic based methods, the One-end Method and Stochastic Propagator Method, for calculating three-point functions in heavy-light semileptonic decays.  The efficiency of these methods was optimized by minimizing the statistical noise, while taking into account the cost incurred by the different set-ups and variance reduction techniques. 

We found that the noise was reduced when placing the stochastic
sources at the sink rather than at the current insertion of the
three-point functions. This enables us to vary the position of the
current but requires that the distance between source and sink is
fixed, very similar to the Sequential Propagator Method.

In terms of noise reduction techniques, full time partitioning was
found to be essential, with no or marginal improvement provided by
additional combinations of color, spin, and even-odd partitioning.

  The One-end Method naively appears promising because of the additional volume average it induces, which also has the effect of projecting out the desired momentum at the  source.  This creates the possibility of reducing the statistical errors, but at a high computational cost.  In our analysis this method clearly fails to outperform the other methods, suggesting it is a poor alternative in heavy-light three-point functions.   

The Stochastic Propagator Method is competitive with the Sequential Propagator Method, and in many cases superior, due to the additional momenta freely available.  The Stochastic Propagator Method is also more versatile, since correlators with different sink smearings, sink interpolating operators, and spectator quark masses can be computed at negligible cost.  This flexibility would allow calculations with the Variational Method and/or partial quenching to be performed with greater efficiency.
 This should further improve with finer lattice spacing, where the momenta  are more finely grained. 

 We have used the Stochastic Propagator Method to calculate the semileptonic form factors at many different values of $q^2$'s, using both the vector current and scalar current.  Our data are parameterized well by the Be\'{c}irevi\'{c} and Kaidalov expressions.  The  results for the $q^2\to 0$ extrapolation, $f_+(0)=0.593(19)$  and $f_+(0)=0.603(24)$, calculated from the vector and scalar current respectively, are comparable to previous determinations.  In the future we will extend this calculation to additional quark masses and lattice spacings, taking advantage of the method's flexibility to improve plateaus and reduce statistical errors.

 \begin{acknowledgments}
 The Chroma software suite \cite{Edwards:2004sx,McClendon} was used extensively in this work.  The simulations were run on the Athene Cluster of the University of Regensburg.  Our work is supported by the DFG Sonderforschungsbereich/Transregio 55 and by the Research Executive Agency of the European Union under grant PITN-GA-2009-238353 (ITN STRONGnet).  Sara Collins acknowledges support from the Claussen-Simon-Foundation (Stifterband f\"ur die Deutsche Wissenschaft).  
 \end{acknowledgments}

\end{document}